\documentclass[a4paper,fleqn,usenatbib]{mnras}

\usepackage[T1]{fontenc}
\usepackage{ae,aecompl}

\usepackage{graphicx}	
\usepackage{amsmath}	
\usepackage{amssymb}	
\usepackage{xspace}
\usepackage{pdflscape}
\usepackage{subfig}
\usepackage{ctable}
\usepackage{threeparttable}
\usepackage{upgreek}

\newcommand{\mcm}{$\upmu$m\xspace}

\title[Star formation in galaxies hosting AGN]{A flat trend of star-formation rate with X-ray luminosity of galaxies hosting AGN in the SCUBA-2 Cosmology Legacy Survey}

\author[J. Ramasawmy et al.]{
Joanna Ramasawmy,$^{1}$\thanks{E-mail: j.ramasawmy@herts.ac.uk}
Jason Stevens,$^{1}$
Garreth Martin$^{1}$
and James E. Geach$^{1}$
\\
$^{1}$Centre for Astrophysics Research, School of Physics, Astronomy and Mathematics, University of Hertfordshire, College Lane, \\ Hatfield AL10 9AB, UK \\
}

\date{Accepted 2019 April 11. Received 2019 April 9; in original form 2019 January 31}

\pubyear{2018}

\begin{document}
\label{firstpage}
\pagerange{\pageref{firstpage}--\pageref{lastpage}}
\maketitle

\begin{abstract}
Feedback processes from active galactic nuclei (AGN) are thought to play a crucial role in regulating star formation in massive galaxies.
Previous studies using \textit{Herschel} have resulted in conflicting conclusions as to whether star formation is quenched, enhanced, or not affected by AGN feedback.
We use new deep 850 \mcm observations from the SCUBA-2 Cosmology Legacy survey (S2CLS) to investigate star formation in a sample of X-ray selected AGN, probing galaxies up to $L_{0.5-7~\rm keV} = 10^{46}$ erg\: s$^{-1}$.
Here we present the results of our analysis on a sample of 1957 galaxies at $ 1 < z < 3 $, using both S2CLS and ancilliary data at seven additional wavelengths (24--500 \mcm) from \textit{Herschel} and \textit{Spitzer}.
We perform a stacking analysis, binning our sample by redshift and X-ray luminosity.
By fitting analytical spectral energy distributions (SEDs) to decompose contributions from cold and warm dust, we estimate star-formation rates for each `average' source.
We find that the average AGN in our sample resides in a star-forming host galaxy, with SFRs ranging from 80--600 $M_{\odot}$ year$^{-1}$.
Within each redshift bin, we see no trend of SFR with X-ray luminosity, instead finding a flat distribution of SFR across $\sim$3 orders of magnitude of AGN luminosity.
By studying instantaneous X-ray luminosities and SFRs, we find no evidence that AGN activity affects star formation in host galaxies.

\end{abstract}




\section{Introduction}

	Attempts to understand the black hole -- galaxy connection have been driven by several key pieces of observational evidence.
	Close correlations are observed in the local Universe between supermassive black hole (SMBH) mass ($M_{\bullet}$) and various properties of the host galaxy bulge, for example: proportionality of $M_{\bullet}$ -- $L_{\mathrm{bulge}}$ \citep{marconi_relation_2003}; $M_{\bullet}$ -- $M_{\mathrm{bulge}}$ \citep{magorrian_demography_1998}; and $M_{\bullet}$ -- $\sigma$, the stellar velocity dispersion of stars in galactic bulges \citep{ferrarese_fundamental_2000, gebhardt_black_2000, haring_black_2004}.
	The existence of these relations forms the basis of an argument for some co-evolutionary process regulating the growth of black holes and their host galaxies: both processes are fueled by an abundance of gas, and this SMBH -- bulge relationship, over spatial scales far exceeding the SMBH's gravitational sphere of influence, suggests a symbiotic evolutionary history.
	The history of SFR density in the universe also corresponds to that of SMBH accretion rate density \citep{boyle_cosmological_1998, franceschini_relationship_1999, aird_evolution_2010}: both processes peak at $z \sim 2$. This observation also supports the hypothesis of a connection between SMBH and its host.

	The mechanism that is most frequently invoked for such a co-evolutionary scenario is an AGN feedback process, which creates a self-regulating system by inhibiting star formation in the host galaxy.
	Feedback mechanisms, such as outflowing winds and jets, exist and have been observed \citep[see review by][]{fabian_observational_2012}, but conclusive signatures of the impact of feedback on the host galaxy are an ongoing observational challenge.

	Cosmological simulations rely on AGN feedback to quench star formation in the most massive galaxies: the bright end of the observed galaxy luminosity function does not match the dark matter halo mass function predicted by $\Lambda$CDM models without invoking AGN feedback \citep[e.g.][]{hopkins_unified_2006, silk_current_2012}.
	Identifying an observational signature of the impact of feedback processes is an important step towards understanding the significance of AGN in galaxy evolution.

	These arguments motivate the large number of studies of star-formation rates in galaxies hosting AGN.
	\citet{page_suppression_2012} investigate the 250 \mcm properties of a sample of X-ray selected AGN over the redshift range $1 < z < 3$. 
	They found that the brightest X-ray sources in the sample were undetected in the submillimetre at the $3 \sigma$ level, interpreting this as evidence of of star formation being suppressed by the most powerful AGN.
	This result prompted a number of responses.
	\citet{barger_host_2015} instead examine individual source properties of X-ray selected AGN from \textit{Chandra} surveys using 850 \mcm data from SCUBA-2 \citep{holland_2013}.
	They find that the mean submillimetre flux density increases with X-ray luminosity to `intermediate' $L_{\mathrm{X}} = 10^{43} - 10^{43.5}$ erg\: s$^{-1}$ and then decreases towards higher luminosities. Using additional \textit{Spitzer} and \textit{Herschel} data they perform individual source SED fitting, finding that FIR luminosities are lower in the highest X-ray luminous sources.
	This is consistent with the first result, and is also interpreted as the suppression of SF in the most X-ray luminous sources. 

	However, the majority of studies find no such signature of SF quenching.
	\citet{harrison_no_2012} perform a stacking analysis of 250 \mcm \textit{Herschel} data, using a larger sample than \citet{page_suppression_2012}.
	They find that while they are able to reproduce Page et al.'s results in the CDFN field, the larger sample shows a constant SFR across the whole X-ray luminosity range, consistent with typical star-forming galaxies at the same redshift.
	\citet{stanley_remarkably_2015} perform an SED fitting analysis similar to \citet{barger_host_2015} incorporating \textit{Spitzer} and \textit{Herschel} data and also find SFRs consistent with the star-forming main sequence of galaxies which do not host AGN.
	Several other studies using \textit{Herschel} data in different fields or redshift ranges \citep{hatziminaoglou_hermes:_2010, shao_star_2010,suh_type_2017} find similar results, as well as a number of studies on optical AGN which find little evidence of AGN-driven quenching \citep{kaviraj_2015, sarzi_2016}.

	A third group of results find increased SF in the most luminous X-ray sources.
	\citet{lutz_laboca_2010} examine the mean stacked 870 \mcm LABOCA fluxes of an X-ray selected sample in the CDFS, dividing their sample into five luminosity bins.
	They find a significant increase in stacked submm flux above $L_{2-10~\rm keV} = 10^{44}$ erg\: s$^{-1}$. Similar results are also reported by \citet{rosario_mean_2012, rovilos_goods-herschel:_2012} and \citet{banerji_cold_2015}.

	Various explanations have been suggested for the disagreement between these studies.
	\citet{harrison_no_2012} suggest that low number statistics and field-to-field variance mean that the results of studies of small (i.e. single field) samples are not reliable.
	Another possibility is that variability in SMBH accretion on much shorter time-scales than that of SF results in instantaneous AGN X-ray luminosities that are not representative of the average black hole accretion rate (BHAR) (e.g. \citealp{hickox_black_2014}).
	Several studies investigate the average X-ray properties of submm detected galaxies \citep{mullaney_hidden_2012,chen_correlation_2013}, finding a positive correlation between SMBH growth and star formation.

	Studies investigating the relationship between star formation and AGN activity typically use \textit{Herschel} data (longest wavelength 500 \mcm) to estimate SFRs.
	More recently, a few studies have used 850 \mcm data from SCUBA-2 \citep{banerji_cold_2015,barger_host_2015} which offers a number of advantages.
	The SCUBA-2 beam is smaller than \textit{Herschel} at its longest wavelengths, resulting in significantly lower confusion noise.
	850 \mcm emission corresponds to rest-frame 212 \mcm emission at z$ = 3$, so the expected contribution from the AGN torus emission is lower than at 500 \mcm (rest frame 125 \mcm at $z = 3$).
	Additionally, over the redshift range of interest in this work ($1 < z < 3$), the negative K-correction \citep{blain_submillimetre_1993} means that the 850 \mcm flux does not diminish with increasing distance for a fixed luminosity, since it is sampling the tail of the dust spectrum.
	Therefore the flux increases with the decreasing rest-frame wavelength that the observed-frame 850 \mcm flux traces with increasing redshift.
	For these reasons, we can more accurately estimate SF at 850 \mcm, and use 850 \mcm flux alone as a probe of SFR.

	These more recent studies incorporating 850 \mcm data have so far only investigated small samples in limited fields.
	Here we present the results of a much larger study at 850 \mcm than has previously been attempted (sample size a factor $\sim 3$ larger than previous work, e.g. \citealp{banerji_cold_2015}), investigating both individual submm detection rates and median stacked `average' sources, over a broader redshift and X-ray luminosity range than equivalent studies using \textit{Herschel} data.
	In section~\ref{sec:Data}, we describe the X-ray, submillimetre and multi-wavelength samples used in this study. Section~\ref{sec:results} presents the detection rates of submillimetre sources, the median stacking procedure adopted for both submillimetre and multi-wavelength data and the analytical SED fitting method used to decompose the FIR spectrum and calculate SFRs. We present the resulting SFRs and investigate the trend with X-ray luminosity. Here we also compare our SFRs with those of a simulated sample of galaxies from the \textsc{Horizon-AGN} simulation \citep{dubois_dancing_2014}. In section~\ref{sec:conclusion}, we summarize our results.

	Throughout this paper we assume a flat cosmology with $H_0 = 69.3$ km s$^{-1}$ Mpc$^{-1}$ and $\Omega_M = 0.287$.

\section{Data}
\label{sec:Data}
	This study combines submillimetre data from the SCUBA-2 Cosmology Legacy Survey (S2CLS, \citealp{geach_scuba-2_2017}) with additional FIR data from \textit{Herschel} SPIRE \citep{griffin_herschel-spire_2010} and PACS \citep{poglitsch_2010}, and \textit{Spitzer} MIPS \citep{rieke_multiband_2004}, to investigate star formation in galaxies hosting AGN. The sample is selected using \textit{Chandra} X-ray surveys; this selection is discussed in section~\ref{ssub:sample}.

	For the initial part of the study, we examine the submillimetre properties of X-ray selected sources in five of the S2CLS survey fields (Akari-NEP, GOODS-N, EGS, COSMOS and UDS), and additionally the Extended \textit{Chandra} Deep Field South (ECDFS) from the LABOCA survey at 870 \mcm \citep{weiss_laboca_2009}.
	The \textit{Chandra} Deep Field South survey is the deepest X-ray survey to date, and so including this field in our sample increases our parameter space to probe the faintest X-ray sources.
	As the S2CLS does not cover the ECDFS, the LABOCA observations provide the highest quality submillimetre survey data in this field.
	This sample of six survey fields will be referred to as the preliminary sample.
	
	The latter part of the study includes additional \textit{Herschel} and \textit{Spitzer} data to investigate the FIR spectrum of sources in the sample across 8 wavelengths between 24 and 850 \mcm.
	We restrict the sample to the COSMOS, GOODS-N and Extended Groth Strip survey fields as these fields have coverage at all 8 wavelengths, however there is the possibility to extend this work in future to include other survey fields which have more limited coverage.
	This sample contains 1397 sources, and will be referred to as the multi-wavelength sample throughout the paper.

	\subsection{X-ray data}
	\label{ssub:sample}
		We identify AGN by their X-ray luminosities, as they are the only compact extragalactic X-ray sources above $L_{2-10~\rm keV} \sim 10^{42} \text{ erg\: s}^{-1}$.
		In the local universe, AGN can be identified by their optical spectra, based on the strength of emission lines (e.g. \citealp{osmer_new_1991}).
		However, this excludes the large fraction of AGN which are optically obscured.
		Objects with radio jets are easily selected by radio surveys, but these constitute only a small fraction of the AGN population.

		Soft X-ray emission is absorbed by cold gas, and sources with column densities $N_{\mathrm{H}} > 10^{24}$ cm$^{-2}$ (`Compton-thick' sources) are obscured in the detectable X-ray regime \citep{risaliti_panchromatic_2004}.
		Thus X-ray selection may exclude the most heavily obscured sources.
		Selecting AGN based on their MIR colours \citep[e.g.][]{stern_mid-infrared_2012} is not biased against obscured galaxies, but colour cuts suffer from contamination by star-forming galaxies, and MIR selected samples of AGN are often incomplete at low luminosity \citep{donley_identifying_2012}.
		X-ray selection remains the method of AGN identification which is most complete and suffers the least contamination. However we note that, by their nature, Compton-thick sources are excluded using this method of identification.

		We use the most recent \textit{Chandra} surveys of the six fields of the preliminary sample, as detailed in Table~\ref{tab:xray_data}.
		These \textit{Chandra} surveys probe deeper (up to 7 Ms / 2 Ms exposures, down to sensitivities of $ 1.9 \times 10 ^{-17}$ / $ 1.2 \times 10 ^{-17}$ erg\: cm$^{-2}$ s$^{-1}$ for the deepest surveys in the preliminary and multi-wavelength samples respectively) than X-ray surveys used in previous work \citep[e.g. $ 9.3 \times 10 ^{-15}$ erg\: cm$^{-2}$ s$^{-1}$, from][]{banerji_cold_2015}.
		By combining both small area, deep surveys and wide shallow surveys, we obtain a well-distributed sample of X-ray luminosities across the redshift range of interest.

	\begin{table*}
		\begin{threeparttable}
			\centering
			\caption{Properties of the S2CLS fields and X-ray catalogues used to create the sample. Columns list the name of the survey field, the area coverage of the S2CLS data, the 1$\sigma$ 850 \mcm depth \citet{geach_scuba-2_2017}, the area coverage of the X-ray data, the exposure time of the X-ray data, the sensitivity of the X-ray data, the number of X-ray detected sources across the whole catalogue, the number of X-ray detected sources within the X-ray luminosity and redshift ranges of interest (see~\ref{ssub:xraylums}). Note that the deep CDFS X-ray survey lies within the sky area of the shallower E-CDFS field, and thus we combine these catalogues; numbers quoted are those in the combined CDFS/E-CDFS field. Submillimetre data for this CDFS/E-CDFS field are from the LABOCA survey at 870 \mcm. Catalogue references are given below the table, including those for spectroscopic redshifts.}
			\label{tab:xray_data}
			\begin{tabular}{l c c c c c c c}
				\hline 
				Field 		& 850 \mcm area	& 1$\sigma$ 850 \mcm depth 	& X-ray area		& X-ray exposure	& X-ray sensitivity	& No. sources \tnote{1}	& No. sources 	\\ 
				\noalign{\smallskip} 
							& (deg$^2$) 		& (mJy beam$^{-1}$ )	& (deg$^2$)			& (ks)				& (erg\: cm$^{-2}$s$^{-1}$) &		& in $L_{\mathrm{X}}$, $z$ \\
				\hline 
				\noalign{\smallskip}
				Akari-NEP 	& 0.6 	& 1.2	&	0.34	&	300		& 	$ 1 \times 10^{-15}$	&	26		& 6		\\ 
				COSMOS			& 2.22	& 1.6	&	2.2		&	160		& 	$ 2 \times 10^{-16}$	&	2287	& 1118	\\
				GOODS-N			& 0.07	& 1.1	&	0.13	&	2000	& 	$ 1.2 \times 10 ^{-17}$	& 	396		& 74	\\
				EGS	 			& 0.32	& 1.2	&	0.29	&	800		& 	$ 3.3 \times 10 ^{-17}$	&	630		& 205	\\
				CDFS 		& 	-	& 	-	&	0.13	&	7000	&	$ 1.9 \times 10 ^{-17}$	&	-		& -		\\
				E-CDFS 			& 0.25	& 1.6	&	0.3		&	228		&	$ 1.9 \times 10 ^{-16}$	&	867		& 220	\\ 
				UDS 	 		& 0.96	& 0.9	& 	1.3		& 	100		&	$ 3.0 \times 10 ^{-15}$	&	586		& 334	\\	 \noalign{\smallskip}\hline
			\end{tabular}
			\begin{tablenotes}
				\item[1] Akari-NEP catalogues from \citet{krumpe_chandra_2015}, \citet{shim_hectospec_2013}; COSMOS from \citet{civano_chandra_2016}, \citet{marchesi_chandra_2016}; GOODS-N from \citet{xue_2_2016}; EGS from \citet{nandra_aegis-x:_2015}; CDFS from \citet{luo_chandra_2016}; E-CDFS from \citet{virani_extended_2006}, \citet{silverman_extended_2010}, \citet{treister_optical_2009}; UDS from \citet{ueda_subaru/xmm-newton_2008}, \citet{akiyama_subaru-xmm-newton_2015}
			\end{tablenotes}
		\end{threeparttable}
	\end{table*}

	\subsection{X-ray luminosities}
	\label{ssub:xraylums}
		Deep optical surveys of these fields result in a large number of sources with accurate spectroscopic redshifts, so we can reject sources with only photometric redshifts and obtain a more reliable sample than used in previous work \citep[e.g.][]{stanley_remarkably_2015,suh_type_2017}.
		For the COSMOS, GOODS-N, EGS and UDS fields, we use spectroscopic redshifts from the catalogs of \citet{marchesi_chandra_2016, xue_2_2016, nandra_aegis-x:_2015, akiyama_subaru-xmm-newton_2015} who match the X-ray sources to optical positions.
		For Akari-NEP and CDFS there are no matched catalogs, so we match to the available spectroscopic catalogs for these fields \citep{shim_hectospec_2013, luo_chandra_2016} using a matching radius of 1.5 arcsec, three times the \textit{Chandra} FWHM of 0.5 arcsec.

		To calculate observed frame fluxes, we use the PIMMS tool \citep{mukai_pimms_1993}.
		Full band 0.5 -- 7 keV source count rates are used to calculate the 0.5 -- 7 keV observed frame fluxes, using a power-law model assuming a photon index of $\Gamma$ = 1.8.
		Galactic hydrogen column densities come from the SWIFT NHtot tool \citep{willingale_calibration_2013}.
		Rest frame X-ray luminosities are then calculated, following the equation:

		\begin{equation}
			L_{0.5-7~\rm keV} = 4 \pi D^{2}_{\mathrm{L}}  F_{0.5-7~\rm keV}  (1 + z)^{\Gamma -2}\
			\label{eq:xray_lum}
		\end{equation}

		\noindent where $D_{\mathrm{L}}$ is the luminosity distance, and $\Gamma$ is the photon index of 1.8.
		X-ray luminosities in the range 0.5 -- 7 keV will hereafter be referred to as $L_{\mathrm{X}}$.\footnote{For comparison with other studies that use 2 -- 10 keV X-ray luminosities, we follow the conversion
		\begin{equation}
			L_{2-10~\rm keV} = 0.721 \times L_{0.5-7~\rm keV}
		\end{equation}
		\citep{xue_2_2016}; see section~\ref{ssub:LAGN}.}

\begin{figure}
			\includegraphics[width=\columnwidth]{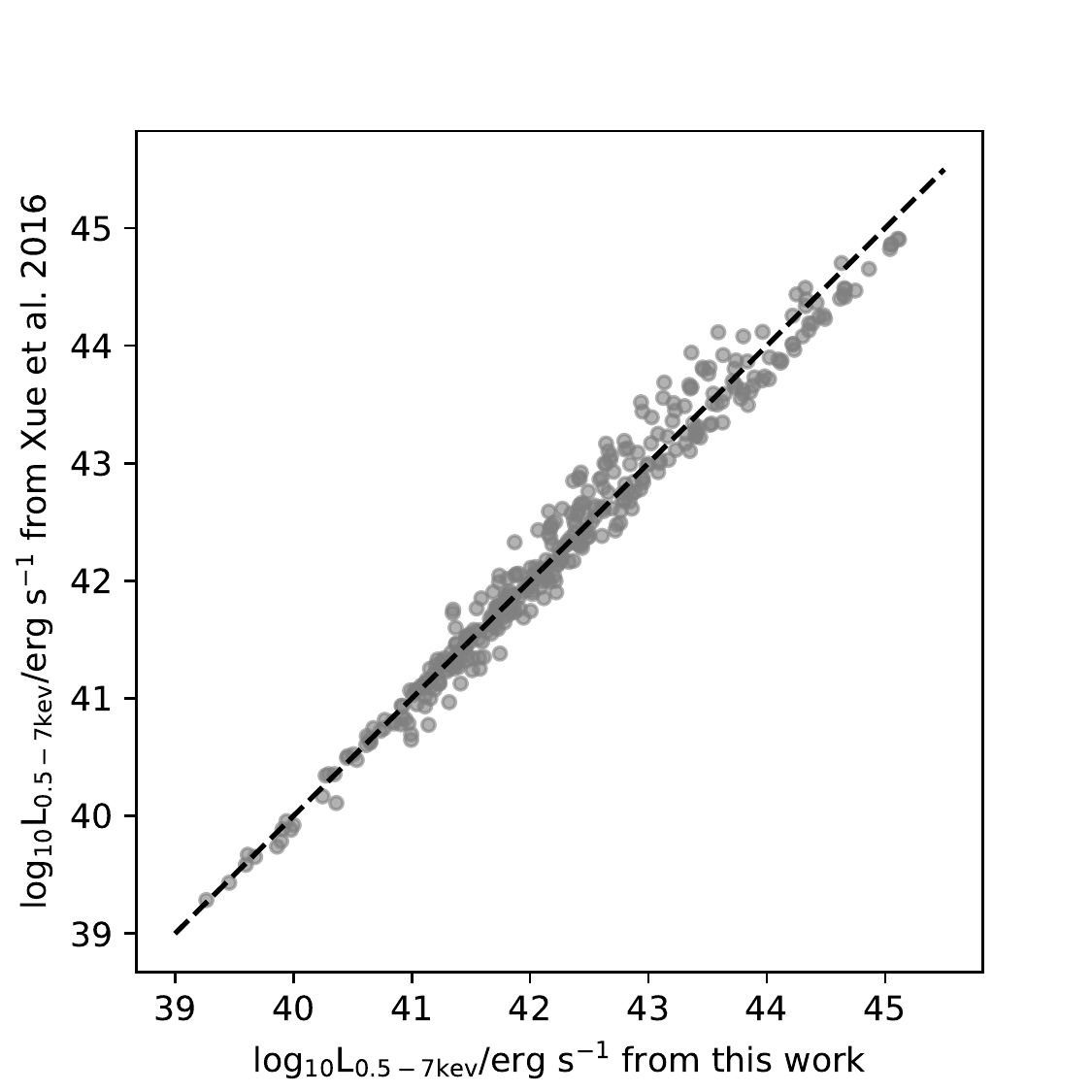}
			\caption{CDFN log $L_{\mathrm{X}}$ values derived in this work, compared with values from \citet{xue_2_2016} which are corrected for intrinsic absorption. The grey dashed line shows 1:1.}
			\label{fig:xlums_compare}
		\end{figure}

		This method does not take into account the intrinsic absorption in obscured sources; however, across the whole sample the X-ray data are not of sufficient quality to allow an in-depth spectral analysis, nor would this significantly affect the luminosities calculated.
		To demonstrate this, we investigate derived X-ray luminosities for sources in the CDFN field, in which the deep 2 Ms survey data allow a more complete analysis. 
		We compare our PIMMS derived X-ray luminosities in this field with those of \citet{xue_2_2016}, who calculate X-ray luminosities using the hard- and soft-band count ratios to correct for intrinsic absorption (Fig.~\ref{fig:xlums_compare}).
		Our simpler method produces $L_{\mathrm{X}}$ values close to one-to-one with the \citet{xue_2_2016} values, with an RMS of 0.18 dex.
		In subsequent analysis, we divide our sample into bins of $\sim 0.8$ dex in log $L_{\mathrm{X}}$, so this level of deviation in our calculated $L_{\mathrm{X}}$ values will have little impact on our results.
		As such this simple method is sufficiently accurate for this work, and allows for consistency across all survey fields in this study.

		\begin{figure*}
			\includegraphics[width=0.9\textwidth]{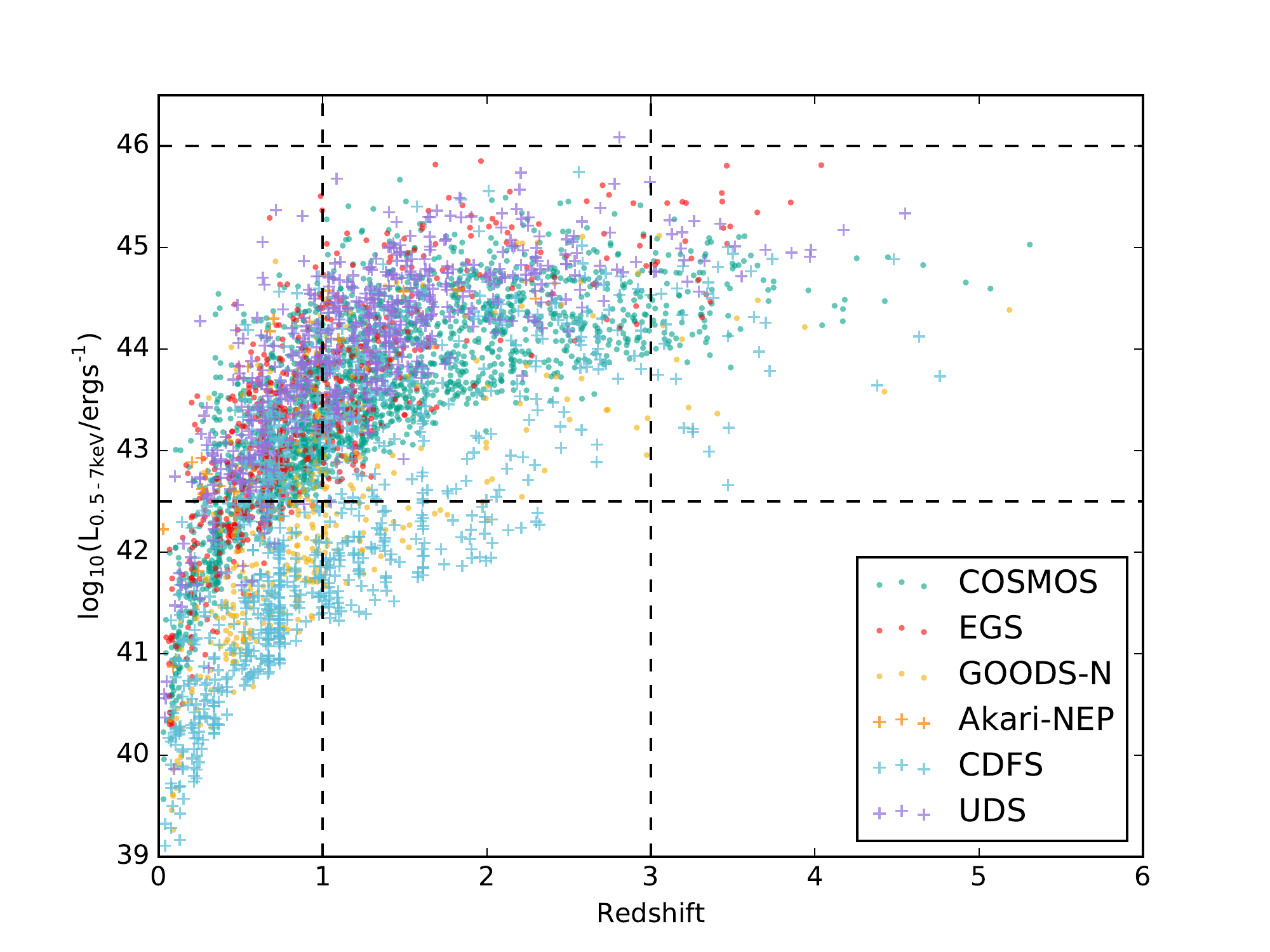}
			\caption{Distribution of sources in redshift and X-ray luminosity. The dashed lines show the ranges of redshift and $L_{\mathrm{X}}$ within which the sample is selected. Round points show sources included in the multi-wavelength sample; crosses are sources that are only in the preliminary sample.}
			\label{fig:lx_z}
		\end{figure*}

		We select sources with a luminosity range of 42.5 < log ($L_{\mathrm{X}}$/erg\: s$^{-1}$ ) < 46, the lower limit chosen to exclude starburst galaxies. 
		We choose a redshift range $1 < z < 3$ to probe the epoch of the peak of SFR density and BH accretion rate density in the universe.
		The distribution of sources in $z$ and $L_{\mathrm{X}}$ space is shown in Fig.~\ref{fig:lx_z}.
		The full sample of all six fields contains 1957 X-ray sources within these redshift and X-ray luminosity ranges.

	\subsection{S2CLS}
	\label{ssub:S2CLS}
		The SCUBA-2 Cosmology Legacy Survey is a large area ($\sim$5 deg$^2$), deep 850 \mcm survey of several well-studied extragalactic fields.
		There are several benefits to using S2CLS to investigate star formation at high redshift rather than previous FIR surveys such as \textit{Herschel}.
		Tracing star formation at long wavelengths (e.g. 850 \mcm with SCUBA-2, compared to 500 \mcm, the longest \textit{Herschel} SPIRE wavelength) probes the tail end of the spectrum of cold dust associated with ongoing star formation, and is less likely to suffer from contamination from AGN-heated dust than shorter wavelengths even at high redshifts. 
		At $z = 3$, observed 850 \mcm corresponds to rest-frame 212 \mcm, while 500 \mcm corresponds to rest-frame 125 \mcm emission in which part of the spectrum an AGN contribution may be significant \citep{symeonidis_agn_2016}. 

		SCUBA-2 also affords improved resolution at long wavelength, with a beam FWHM of 14.8 arcsec, compared to the \textit{Herschel} SPIRE beam FWHM of 36.4 arcsec at 500 \mcm \citep{griffin_herschel-spire_2010}.

		Additionally, over the redshift range used in this study ($1 < z < 3$), the negative K-correction \citep{blain_submillimetre_1993} results in 850 \mcm flux that does not diminish with redshift for a given luminosity.
		Thus a simple comparison of fluxes may be made as a preliminary analysis (see section~\ref{ssub:stacking}).

	\subsection{LABOCA data}
	\label{ssub:laboca}
		We also use submm data from the LABOCA survey of the ECDFS at 870 \mcm \citep{weiss_laboca_2009}, as the CDFS X-ray survey is the deepest available X-ray survey data, with an exposure time of 7 Ms.
		While the LABOCA data are of lower resolution, with a beam width of 19.2 arcsec, the are the highest quality submm data available for the CDFS, and including this field in our sample allows us to probe the low X-ray luminosity, high $z$ parameter space that would otherwise be excluded.
		The LABOCA ECDFS catalogue contains 126 submm detected sources at $> 3.7 \sigma $.

	\subsection{FIR data}
	\label{ssub:FIR}

		To investigate the FIR spectrum of our sources, we take advantage of the wealth of \textit{Herschel} and \textit{Spitzer} data available.

		\textit{Herschel}-SPIRE maps at 500, 350 and 250 \mcm come from the HerMES survey \citep{hermes_collaboration_herschel_2012}.
		SPIRE maps are in units of Jy/beam, and so fluxes are extracted from the central pixel of stacked images (see section~\ref{ssub:stacking} for further details on the stacking procedure).

		We use combined \textit{Herschel}-PACS data from the PEP \citet{lutz_pacs_2011} and GOODS-Herschel \citet{elbaz_goods-herschel:_2011} programs, as described in \citet{magnelli_deepest_2013}. To extract 100 and 160 \mcm fluxes from the PACS maps, we perform aperture photometry following the guidelines in the PEP DR1 readme; for the 100 \mcm and 160 \mcm maps, we use aperture radii of 7.2 and 12 arcsec, respectively.

		We also use \textit{Spitzer}-MIPS \citep{rieke_multiband_2004} maps at 24 and 70 \mcm from the S-COSMOS \citep{sanders_s-cosmos:_2007} and FIDEL \citep{dickinson_deep-wide_2006} surveys. 24 \mcm data for the GOODS-N field are from the GOODS survey \citep{dickinson_great_2003}.
		Again we perform aperture photometry on the MIPS maps. 
		By comparing extracted fluxes with those of catalog sources in these maps, we determine the most accurate aperture radii, and use radii of 3.5 and 16 arcsec at 24 and 70 \mcm respectively.

		For each of the 8 FIR and submillimetre wavelengths, we verify that there is no significant systematic positional offset ( $\leq 1$ pixel) by performing a median stack at all X-ray source positions and examining the centroids of the resulting images.

\section{Results and discussion}
\label{sec:results}

	\subsection{Detection rates of submillimetre sources}
	\label{ssub:detections}
		We first explore the fraction of sources that are detected in the S2CLS catalogues.
		The catalogues use a detection threshold of 3.5$\sigma$ \citep{geach_scuba-2_2017}.
		As the S2CLS survey fields are of slightly different depths, we impose a lower threshold on noise of 1.2 mJy beam$^{-1}$, the average 1$\sigma$ depth of S2CLS, across each field to exclude sources that are detected below this noise level in the deepest fields.
		To match sources from the submillimetre catalogues to the X-ray positions, we use a matching radius of 4$\sigma_{\mathrm{pos}}$, where $\sigma_{\mathrm{pos}}$ is the positional uncertainty given by the equation:
		
		\begin{equation}
			\sigma_{\mathrm{pos}} = \frac {0.6 \times \mathrm{FWHM}}{\mathrm{SNR}}
		\end{equation}

		\noindent \citep{ivison_scuba_2007}.

		This results in a mean matching radius of $\sim 7$ arcsec. The \textit{Chandra} positions are accurate to $\approx 0.5$ arcsec, so we neglect any X-ray positional uncertainty as it is an order of magnitude smaller than the submillimetre positional accuracy.
		In the case that multiple sources are within the matching radius, we take the closest match.
		We test the spurious detection rate by matching the S2CLS catalogue positions with a randomly generated set of coordinates of the same size as the source list, repeating this procedure 1,000 times.
		This results in a spurious matching fraction of 1 per cent.
		We also include the ECDFS field using the LABOCA catalogues (see section~\ref{ssub:laboca}).

\begin{figure}
			\includegraphics[width=\columnwidth]{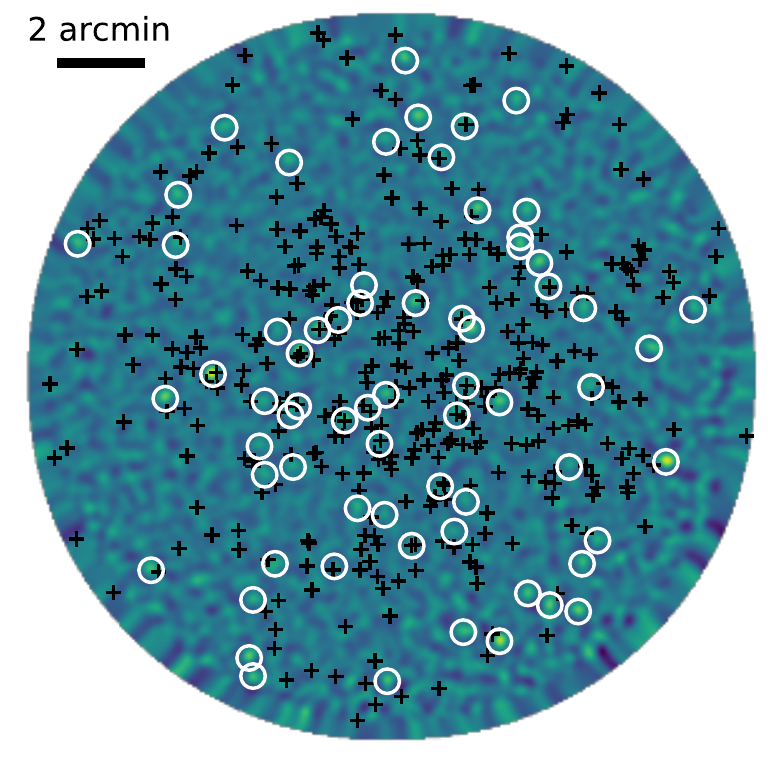}
			\caption{The S2CLS map of the GOODS-N field. Submillimetre detections from the S2CLS catalogue (\textgreater 3.5 $\sigma$) are marked with white circles, and X-ray detections are marked with black crosses. Where there are two X-ray sources within the matching radius of an S2CLS source, the nearest match is assumed to be the correct counterpart. The majority of X-ray sources are not detected at 850 \mcm at the 3.5 $\sigma$ level.}
			\label{fig:goodsN}
		\end{figure}

		Using this matching criterion, we find that 39 sources are detected at 850  \mcm, 2 per cent of the whole sample.
		Fig.~\ref{fig:goodsN} shows an example S2CLS field (GOODS-N), with submillimetre detections marked with white circles, and X-ray detections marked with black crosses.

\begin{figure}
			\includegraphics[width=\columnwidth]{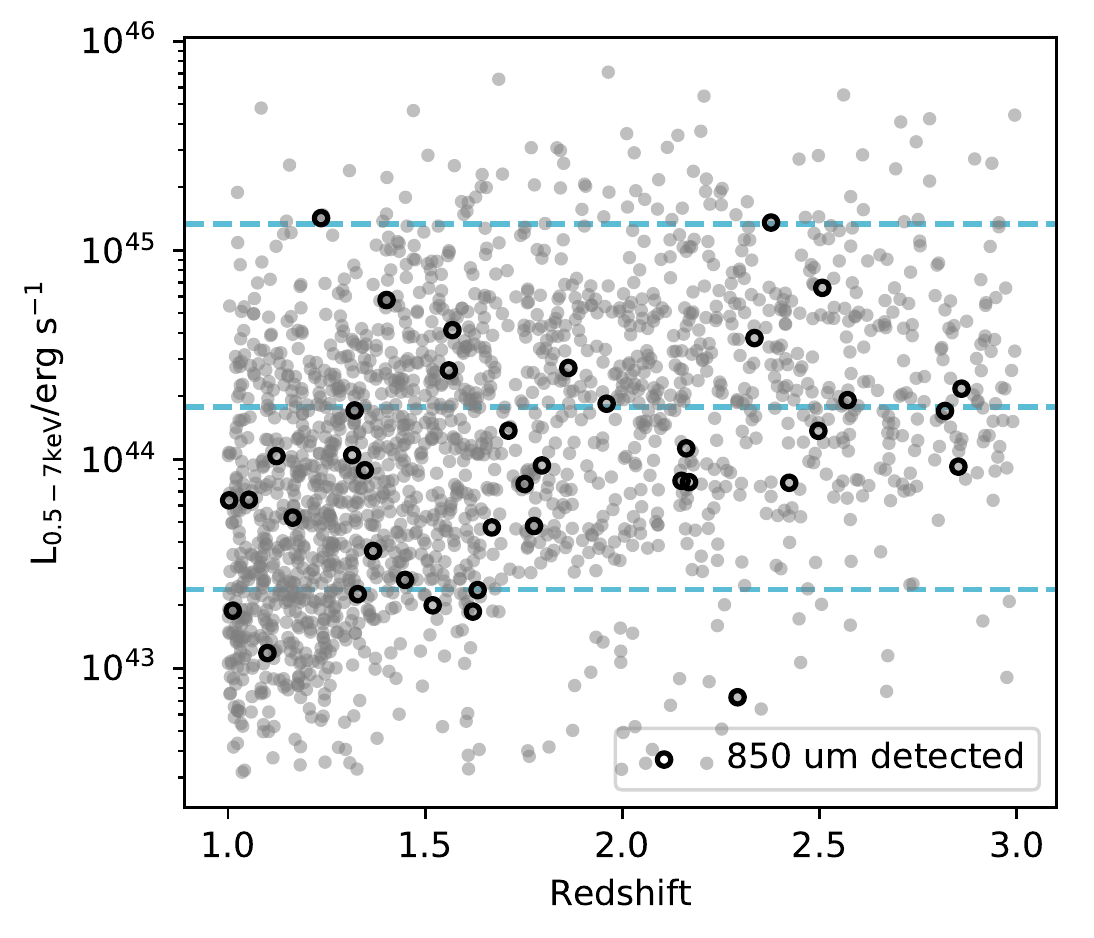}
			\caption{Sources detected (black open circles) or undetected (grey circles) at 850 \mcm across all fields, in z/$L_{\mathrm{X}}$ space. The boundaries marking each of the four $L_{\mathrm{X}}$ bins are marked by dashed lines. }
			\label{fig:matching}
		\end{figure}

		We split the sample into four $L_{\mathrm{X}}$ bins, spaced equally in the log $L_{\mathrm{X}}$ range of the sample, and investigate whether the detection fraction varies with X-ray luminosity.
		Fig.~\ref{fig:matching} shows the positions of the submillimetre detected sources in redshift and $L_{\mathrm{X}}$ space.
		We construct a contingency table of the numbers of detections and non-detections in each $L_{\mathrm{X}}$ bin, shown in Table~\ref{tab:detections}.
		In contrast to \citet{page_suppression_2012}, we find similar detection fractions across all X-ray luminosities; the highest luminosity bin has the highest detection fraction (2.53 per cent) but also contains the smallest number of sources.
		We perform a $\chi^{2}$ test of independence on the contingency table, to investigate whether the difference in detection fractions is significant betweeen the luminosity bins.
		Assuming that there is no trend with $L_{\mathrm{X}}$, this results in a $p$-value of 0.43, thus we cannot reject the null hypothesis that the detection fraction is independent of $L_{\mathrm{X}}$.
		Moreover, a $\chi^{2}$ test for trend across the $L_{\mathrm{X}}$ bins results in a $p$-value of 0.83, thus we cannot reject the null hypothesis that there is no trend in the number of detections with $L_{\mathrm{X}}$.

		While the detection fraction at 850 \mcm is very low (2 per cent), it is double the spurious detection rate, implying that we do see some real submillimetre sources associated with X-ray positions.
		However, it is worth noting that this kind of source-by-source analysis is severely limited by the quality of the data, and alone cannot be used to draw a conclusion that is statistically robust.
		This motivates the use of a stacking analysis (see section~\ref{ssub:850_stack}), which, although coarser, is a useful method given the challenge that these data pose for individual source statistics.

		\begin{table}
			\centering
			\caption{Number of X-ray sources, and those with submillimetre detections, in each $L_{\mathrm{X}}$ bin. Column 1 shows the range in $L_{\mathrm{X}}$ of each bin, column 2 shows the number of sources, column 3 shows the number of 850 \mcm detections, column 4 shows the detection fraction as a percentage, and column 5 shows the median redshift in each bin.}
			\label{tab:detections}
			\begin{tabular}{lcccr} 
				\hline
				log $L_{\mathrm{X}}$(erg\: s$^{-1}$)& no. sources & 850 \mcm ? & \% & $z_{\mathrm{m}}$\\
				\hline
				42.5 -- 43.4  & 321 & 7   & 2.18 & 1.19 \\
				43.4 -- 44.3 	& 865 & 21  & 2.43 & 1.45 \\
				44.3 -- 45.1	& 692 & 9   & 1.30 & 1.80 \\
				45.1 -- 46 	& 79  & 2   & 2.53 & 2.05 \\
				\hline
			\end{tabular}
		\end{table}

		\subsubsection{Redshift dependence}

\begin{figure}
			\includegraphics[width=\columnwidth]{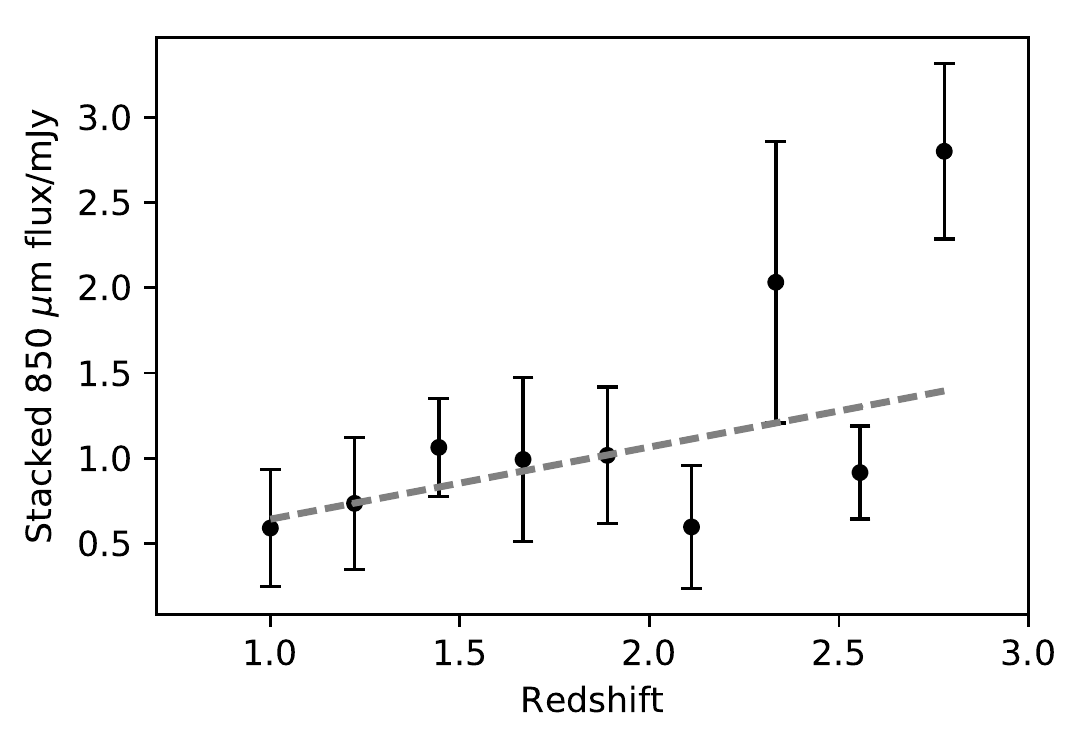}
			\caption{Stacked 850 \mcm flux with redshift. The dashed line shows a weighted least squares fit to the data points, which gives a slope of 0.42; errorbars show 1-$\sigma$ uncertainties on the stacked fluxes.}
			\label{fig:redshift_dependence}
		\end{figure}

		In this analysis, we have not taken into account any selection effects due to the redshift distribution of sources; the highest luminosity bin has a higher median redshift ($z = 2.05$) than the lowest bin ($z = 1.19$), see Table~\ref{tab:detections}.
		Any relationship therefore between redshift and 850 \mcm flux is neglected.
		However, a correlation is to be expected given the increasing number density of submillimetre sources from $1 < z < 3$ \citep[e.g.][]{chapman_redshift_2005}, and the redshift evolution of SFR \citep[e.g.][]{blain_history_1999}.
		In Fig.~\ref{fig:redshift_dependence} we show the stacked submillimetre flux binned by redshift, and indeed there is a positive correlation between submillimetre flux and redshift.
		The Pearson coefficient for these data is 0.67; a weighted least squares fit gives a slope of 0.42. 

\begin{figure*}
			\includegraphics[width=0.7\textwidth]{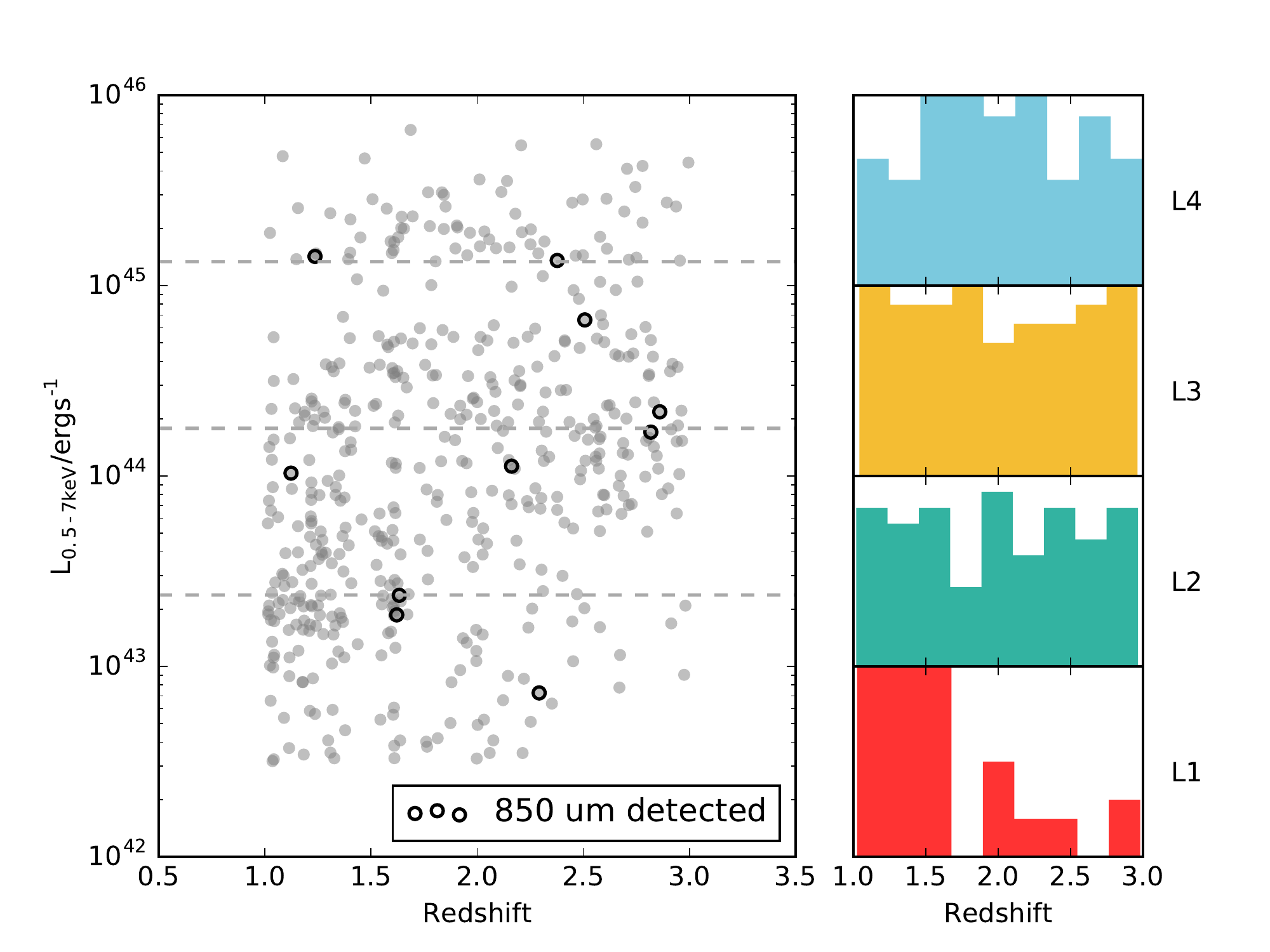}
			\caption{Left: Sources detected or undetected in submm in z/$L_{\mathrm{X}}$ space for the redshift matched sample. The boundaries marking each of the four $L_{\mathrm{X}}$ bins are marked by grey dashed lines; bin widths are defined by splitting the sample range evenly in log space. Right: Distributions in redshift in each of the 4 $L_{\mathrm{X}}$ bins.}
			\label{fig:matching_even}
		\end{figure*}

		To account for this, we perform the same analysis on a matched sample that is evenly distributed in redshift space across each of the $L_{\mathrm{X}}$ bins.
		The distribution of redshifts in each bin of the matched sample is shown in the right panel of Fig.~\ref{fig:matching_even}.
		With the exception of the lowest luminosity bin, which simply does not have enough high redshift sources to create a significant evenly distributed sample, the distributions are matched in their redshift distributions.
		A two sample Kolmogarov-Smirnov test, testing each bin against the 3rd luminosity bin, as an example, shows that we cannot reject the null hypothesis that each sample is drawn from the same distribution. 
		The left panel of Fig.~\ref{fig:matching_even} shows the results of matching detections in this evenly distributed sample.
		Again, we find a higher detection fraction ($\sim$ 2.8 per cent) in the highest luminosity bin; again this bin has the smallest sample size.
		As in section~\ref{ssub:detections},we perform the same $\chi^{2}$ test for independence on the detection fractions across the four bins.
		Assuming no trend with $L_{\mathrm{X}}$, we obtain a $p$-value of 0.89.
		Testing for a trend with $L_{\mathrm{X}}$, we obtain a $p$-value of 0.99.
		Neither allow use to reject the null hypothesis that the detection fraction is independent of $L_{\mathrm{X}}$.
		Accounting for the redshift distribution of the sources, we find no significant difference in detection rates across the $L_{\mathrm{X}}$ bins (see Table~\ref{tab:detections_even}).

		\begin{table}
			\centering
			\caption{Number of X-ray sources, and those with submillimetre detections, in each $L_{\mathrm{X}}$ bin, for the redshift matched sample. Column labelling follows that of Table~\ref{tab:detections}.}
			\label{tab:detections_even}
			\begin{tabular}{lcccc} 
				\hline
				log $L_{\mathrm{X}},z$(erg\: s$^{-1}$) & no. sources & 850 \mcm ? & \% & $z_{\mathrm{m}}$\\
				\hline
				42.5 -- 43.4  & 116 & 3  & 2.59 & 1.52 \\
				43.4 -- 44.3 	& 169 & 3  & 1.78 & 2.00 \\
				44.3 -- 45.1	& 131 & 7  & 2.59 & 1.99 \\
				45.1 -- 46 	& 71  & 5  & 2.82 & 2.05 \\
				\hline
			\end{tabular}
		\end{table}
	\subsection{850 \mcm stacked fluxes}
	\label{ssub:850_stack}

		As the majority of sources are undetected at submm wavelengths, we divide the sample into bins of X-ray luminosity and stack fluxes in each bin to measure an average flux density.

\begin{figure}
			\includegraphics[width=\columnwidth]{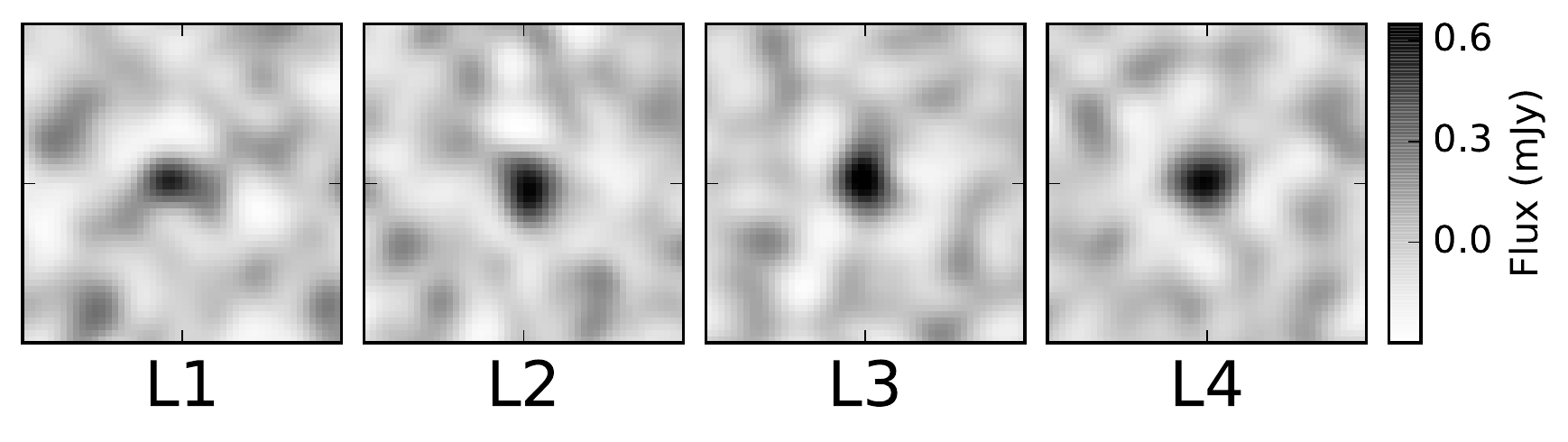}
			\caption{Median stacked images for each $L_{\mathrm{X}}$ bin (see Table~\ref{tab:850_stacks}). Each image is 100 $\times$ 100 arcsec.}
			\label{fig:850_stacks}
		\end{figure}

		We begin by investigating the 850 \mcm fluxes.
		Sources are divided into four bins of X-ray luminosity.
		For comparison with the latter part of the study, we use the same $L_{\mathrm{X}}$ limits to define bins as in section~\ref{ssub:stacking}, resulting in similar numbers of sources in each bin (see Table~\ref{tab:850_stacks}). 
		We stack at every X-ray position, whether or not the source is detected at 850 \mcm.
		100 $\times$ 100 arcsec images are cut around each X-ray position.
		We begin by calculating both a linear mean and median stack, by calculating the mean/median pixel values.
		An `average' image is therefore created, for each bin in $L_{\mathrm{X}}$.
		Examples of median stacked images in each bin are shown in Fig.~\ref{fig:850_stacks}.

		\renewcommand{\arraystretch}{1.3}
		\begin{table*}
			\centering
			\caption{Stacked 850 \mcm flux in each of four $L_{\mathrm{X}}$ bins. Column 1 shows the name of the bin, column 2 the range of X-ray luminosities, column 3 shows the number of sources in each bin and column 4 shows the 850 \mcm median stacked flux in mJy.}
			\label{tab:850_stacks}
			\begin{tabular}{ccccc} 
				\hline
				$L_{\mathrm{X}}$ bin & $L_{\mathrm{X}}$ range ($10^{43}$ erg\: s$^{-1}$ )& no. sources & median flux (mJy) & mean flux (mJy)\\
				\hline
				L1	&	0.351 -- 3.48	& 359	& $ 0.30 _{- 0.12} ^{+ 0.22 } $ & $ 0.45 _{- 0.04} ^{+ 0.11}$\\
				L2 	&	3.48 -- 9.55 	& 389	& $ 0.53 _{- 0.30} ^{+ 0.24 } $ & $ 0.63 _{- 0.04} ^{+ 0.11}$\\
				L3 	&	9.55 -- 26.6 	& 414	& $ 0.50 _{- 0.19} ^{+ 0.23 } $ & $ 0.67 _{- 0.04} ^{+ 0.11}$\\
				L4 	&	26.6 -- 710. 	& 474	& $ 0.48 _{- 0.15} ^{+ 0.22 } $ & $ 0.61 _{- 0.03} ^{+ 0.09}$\\
				\hline
			\end{tabular}
		\end{table*}
		\renewcommand{\arraystretch}{1}

\begin{figure}
			\includegraphics[width=\columnwidth]{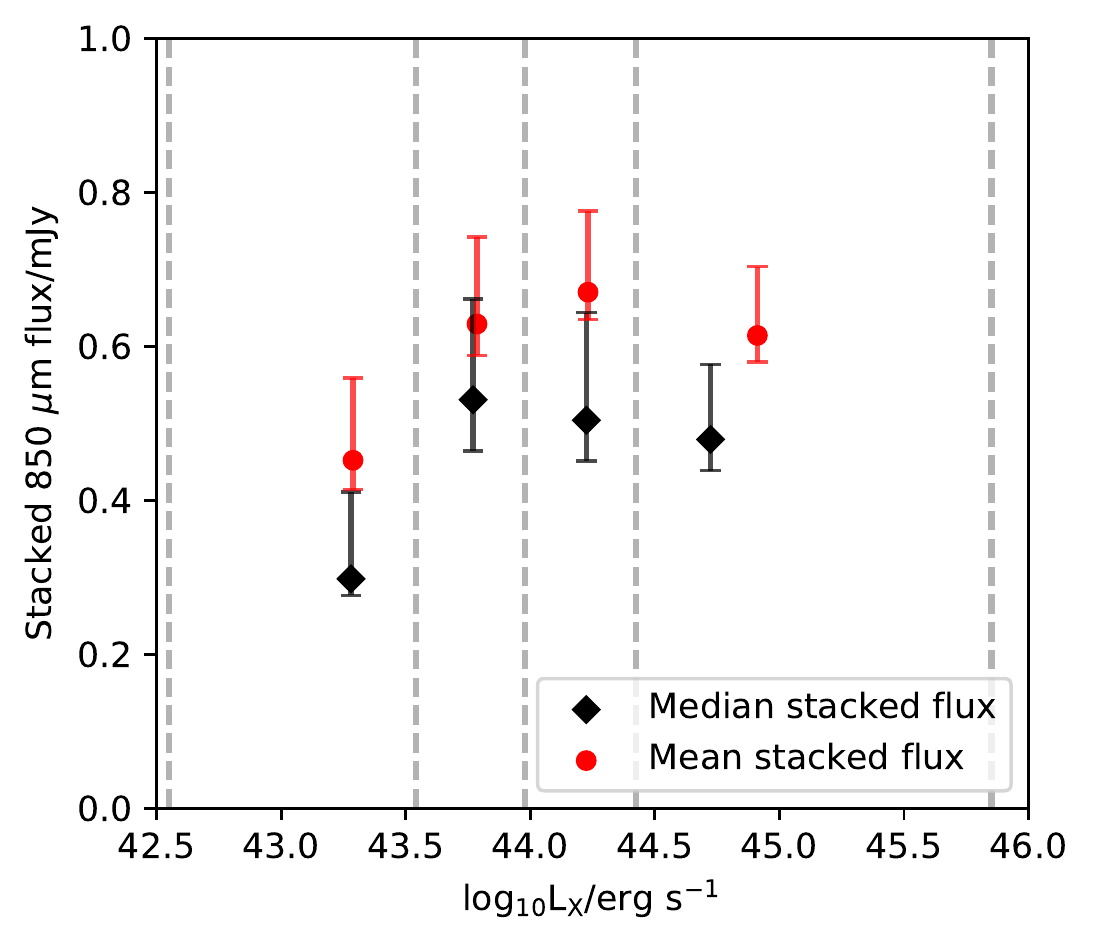}
			\caption{Mean (red circles) and median (black diamonds) stacked 850 \mcm flux with $L_{\mathrm{X}}$. Points are plotted at the mean/median $L_{\mathrm{X}}$ of each bin respectively. Grey dashed lines show the $L_{\mathrm{X}}$ limits of each bin. Error bars show the 1$\sigma$ uncertainties calculated with a bootstrapping analysis.}
			\label{fig:850_stacks_plot}
		\end{figure}

		Fig.~\ref{fig:850_stacks_plot} shows the mean and median stacked fluxes read from the central pixel of each of these images.
		Uncertainties on these fluxes are calculated via a bootstrap analysis.
		For each bin, we create a new stack by drawing the same number of images as in the bin randomly and with replacement from the original stack.
		We then calculate the mean and median stacked fluxes in these new stacks.
		Repeating this 10,000 times, a distribution of the means/medians is produced, from which we take the 36th and 84th percentiles as the 68 per cent confidence interval, taking this to be equivalent to the 1$\sigma$ uncertainty.

		The mean stacked fluxes are strongly biased to higher stacked fluxes by the few very bright submillimetre sources in each stack \citep[e.g.][]{barger_host_2015}.
		The median stacks are therefore a more accurate estimator of the average properties of sources binned by $L_{\mathrm{X}}$, robust to bright outliers, and thus we choose to use the median stacks throughout the rest of this work.
		However, both the mean and median stacked fluxes show the same trend across the $L_{\mathrm{X}}$ bins, and so this choice does not affect the conclusion that we draw from our results.
		We do not observe a decrease in flux with increasing $L_{\mathrm{X}}$.
		Instead, the stacked fluxes are consistent with a flat trend within the 68 per cent confidence intervals shown in the highest three bins.
		Assuming that $L_{\mathrm{X}}$ accurately traces AGN activity, and that 850 \mcm flux is a good tracer of star formation, this result may be interpreted as no evidence that AGN activity suppresses star formation in the galaxies in our sample.

		However, this stacked 850 \mcm flux approach is a simplistic one, and neglects some potentially important factors, namely:

		\noindent 1) We assume that 850 \mcm flux traces star formation alone, and do not account for any contribution to the flux from the AGN torus.
		While this may be a valid assumption at low $z$, at the high $z$ end of our sample ($z = 3$) observed 850 \mcm emission corresponds to 212 \mcm rest-frame emission, which may be affected by a contribution to the flux from the AGN torus \citep[e.g.][]{symeonidis_agn_2016}. 

		\noindent 2) The mean redshift of sources in each of the bins changes, with the highest $L_{\mathrm{X}}$ bin containing the most high redshift sources.
		As such, the part of the SED being sampled at 850 \mcm changes, and this is not taken into account in such a stacking analysis.

		To determine a more accurate measure of star-formation rate, we require a method which separates the contributions to the FIR and submm spectrum from the cold dust associated with ongoing star formation, and that from the AGN torus.

	\subsection{Multi-wavelength stacked fluxes}
	\label{ssub:stacking}
		In order to investigate the FIR/submm spectra of sources in the multi-wavelength sample, we median stack at each of the 8 wavelengths to create average sources to which we can fit SEDs (section~\ref{ssub:SED}).
		The sample is first divided into four redshift bins, so that the stacked fluxes can be used for SED fitting without the K-correction becoming significant within each bin; the maximum width of redshift bin is from $2.05 < z < 3$, which for observed 850 \mcm fluxes corresponds to a range in rest-frame wavelengths of 65 \mcm.
		These redshift bins are then further subdivided by $L_{\mathrm{X}}$ into four evenly populated bins.
		This results in 16 bins, each containing $\sim$90 sources (see Table~\ref{tab:bins}).

		\renewcommand{\arraystretch}{1.3}
		\begin{table}
			\centering
			\caption{Median properties of each of the 16 bins in redshift and X-ray luminosity. Column 1 shows the redshift range of each bin, column 2 shows the number of sources, column 3 the median redshift, column 4 the median X-ray luminosity in erg\: s$^{-1}$, and column 5 the calculated SFR in solar masses per year, as described in section~\ref{ssub:LAGN}.}
			\label{tab:bins}
			\begin{tabular}{ccccc}
				\hline
				z-range & No.  & $z_{m}$ & $L_{\mathrm{X}}$ & SFR  \\
				& sources &  & ($10^{43}$ erg\: s$^{-1}$ ) & ($M_{\odot}$ yr$^{-1}$) \\
				 \hline
				1 -- 1.24 & 93 &  1.10  & $ 1.12 $ & $85   _{- 5   } ^{+  8   }$ \\   
				& 92 &  1.10  & $ 2.55 $ & $77   _{- 12  } ^{+  5   }$ \\  
				& 92 &  1.14  & $ 5.95 $ & $79   _{- 5   } ^{+  9   }$ \\  
				& 93 &  1.09  & $ 20.7 $ & $100  _{- 14  } ^{+  7   }$ \\
				 \hline 
				1.24 -- 1.53 & 89 &  1.32  & $ 1.87 $ & $144  _{- 25  } ^{+  18  } $ \\ 
				& 88 &  1.38  & $ 3.84 $ & $122  _{- 22  } ^{+  24  } $ \\   
				& 88 &  1.33  & $ 10.5 $ & $167  _{- 28  } ^{+  11  } $ \\  
				& 89 &  1.40  & $ 31.7 $ & $137  _{- 12  } ^{+  31  } $ \\
				 \hline 
				1.53 -- 2.05 & 87 &  1.76  & $ 3.70 $ & $155  _{- 23  } ^{+  18  } $ \\ 
				& 86 &  1.76  & $ 7.37 $ & $299  _{- 43  } ^{+  18  } $ \\ 
				& 86 &  1.80  & $ 19.3 $ & $351  _{- 64  } ^{+  29  } $ \\  
				& 87 &  1.88  & $ 50.4 $ & $366  _{- 37  } ^{+  49  } $ \\
				 \hline
				2.05 -- 3 & 87 &  2.47  & $ 7.46 $ & $592  _{- 50  } ^{+  95  } $ \\  
				& 86 &  2.53  & $ 15.6 $ & $600  _{- 62  } ^{+  141 } $ \\ 
				& 86 &  2.39  & $ 30.0 $ & $453  _{- 157 } ^{+  71  } $ \\ 
				& 87 &  2.45  & $ 110  $ & $588  _{- 54  } ^{+  114 } $ \\
				 \hline
			\end{tabular}
		\end{table}
		\renewcommand{\arraystretch}{1}

		Following the method outlined in section~\ref{ssub:850_stack}, we create median stacked images for each wavelength in each bin, then measure fluxes from these images.
		For the S2CLS and SPIRE images, the flux is measured at the central pixel of the stacked image. For the MIPS and PACS images, we perform aperture photometry on the stacked images, as described in section~\ref{ssub:FIR}.

		At SPIRE wavelengths, there will be a significant contribution to the flux from confusion noise, which may bias our results.
		We perform a bootstrapping method to estimate the confusion noise in each bin at each wavelength to characterise the noise and to account for this bias.
		To do so, we bootstrap over stacks at random positions in the survey images.
		In each bin, we draw an equal number of random positions as the number of sources in the bin, and repeat this random stack 10,000 times.
		We measure the stacked flux from these random stacks, and use the resulting distribution to characterise the noise.
		We find that the random stacks are not centered around zero, but rather are offset by some positive flux.
		This is likely due to blending of faint sources, and as such some positive flux density accumulates in the stacked image.
		This procedure is performed for each bin at each wavelength.
		We measure the offsets for each, and subtract them from the measured stacked fluxes, following the procedure of \citet{stanley_mean_2017}.

		A similar bootstrap procedure is performed to estimate the 1$\sigma$ error on the stacked median fluxes.
		Again, as each bin contains a different number of sources, we perform this procedure on each bin in each wavelength separately.
		An equal number of sources as the number of sources in each bin are drawn randomly with replacement from the sample, then the median stacked image is created and the flux measured.
		We do this 10,000 times to create a distribution of median fluxes, from which we take the 68 per cent confidence interval as a 1$\sigma$ error estimate.
		The resulting distributions are smooth and gaussian-like for those wavelengths for which we perform aperture photometry (160, 100, 70 and 24 \mcm), but for the wavelengths in which we simply measure the central pixel flux, the possible values for the medians are discrete rather than continuous and as a result the distribution is not smooth.
		To account for this, we perform a smoothed bootstrap on these four wavelength samples (850, 500, 350 and 250 \mcm), in which we add a small amount of random noise ($\sigma = 0.1$) to each measured median stacked flux.
		This smoothed bootstrap distribution is then smooth and gaussian-like, and representative of the spread of the sample.
		From this we measure the 68 per cent confidence interval as the 1$\sigma$ error.

		\begin{figure}
			\includegraphics[width=\columnwidth]{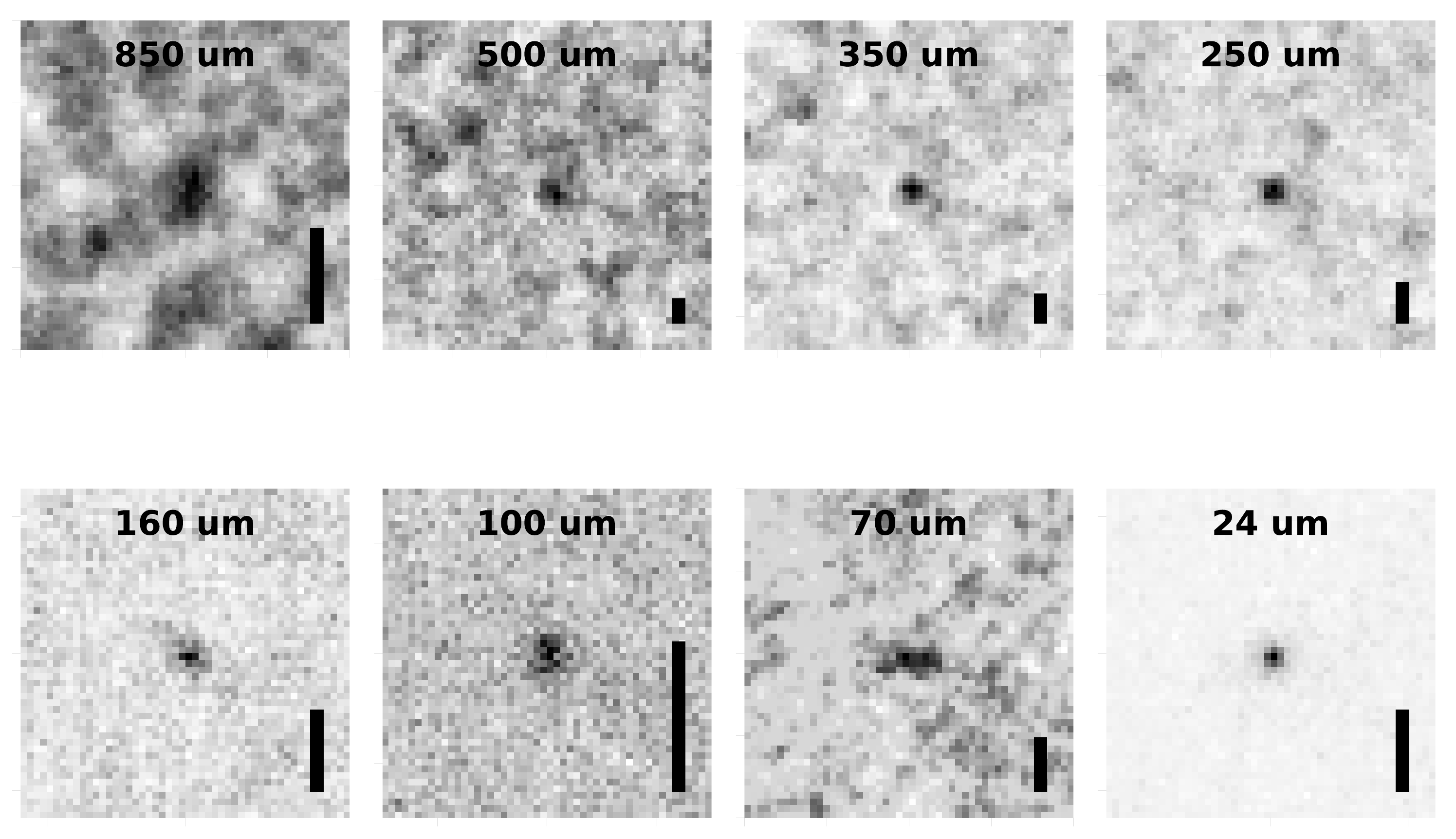}
			\caption{Example stacked images in each wavelength. The black bar on each image indicates 25 arcsec scale.}
			\label{fig:stacks}
		\end{figure}

		This process results in measured median stacked fluxes for each of the 8 wavelengths, binned by $L_{\mathrm{X}}$  and redshift, creating 16 `average' sources spanning the ranges of $z$ and $L_{\mathrm{X}}$ of interest.

	\subsection{SED fitting}
	\label{ssub:SED}

		Using the resulting fluxes, we fit an SED to each of the 16 `average' sources.
		We model the FIR SED as two components: a cold dust greybody contribution, and a warm dust power law contribution, described by the following equation:

		\begin{equation}
			S(\lambda) = N_{\mathrm{gb}} \frac{(\frac{c}{\lambda})^{\beta + 3}}{e^{\frac{h c}{\lambda k T}} - 1} + N_{\mathrm{pl}} \lambda^{\alpha} e^{-\frac{\lambda}{\lambda_{\mathrm{c}}}^{2}}
		\end{equation}		

		\noindent simplified assuming the optically thin case, where $S(\lambda)$ is the flux in Jy, $N_{\mathrm{gb}}$ and $N_{\mathrm{pl}}$ are a normalization coefficients of the greybody and power law components respectively, $T$ is the temperature of the greybody, $\beta$ is the emissivity index of the dust grains, $\alpha$ is the slope of the power law, and $\lambda_{\mathrm{c}}$ is the cut-off wavelength of the power law \citep{casey_far-infrared_2012}.
		We fix $\beta$ to be 1.5, a value widely assumed in the literature \citep[e.g.][]{da_cunha_simple_2008,casey_far-infrared_2012}.
		Following arguments in \citet{smith_isothermal_2013}, allowing $\beta$ to be a free parameter would have little impact on our derived FIR luminosities.
		This leaves five free parameters for the fit: $N_{\mathrm{gb}}$, $N_{\mathrm{pl}}$, $T$, $\alpha$ and $\lambda_{\mathrm{c}}$.

		Fitting an analytical form of the SED, rather than using a library of templates, is a sensible choice here as (a) the data are not sufficiently high-resolution to warrant fitting to the level of detail of empirical templates, and (b) this removes biases introduced by choice of templates, which include restricting the AGN contribution to the spectrum (see below), and allows us to test a more flexible model.
		Our results are in good agreement with studies such as \citet{stanley_remarkably_2015} and \citet{suh_type_2017} that use a template fitting approach (see section~\ref{ssub:LAGN} and Fig.~\ref{fig:Lir_Lagn}), suggesting that template fitting is not necessary to decompose the FIR spectrum.

		We choose to model a single warm dust component with a power law with exponential cutoff, and a single cold dust component with the greybody function, and do not include any additional component representative of kpc-scale dust heating by the AGN \citep{symeonidis_agn_2016}.
		However, in contrast to previous analytical approaches \citep[e.g.][]{barger_host_2015}, we do not fix the cut-off wavelength of the power law component $\lambda_{\mathrm{c}}$, allowing this as a free parameter.
		Fixing $\lambda_{\mathrm{c}}$ to a given wavelength risks underestimating the AGN contribution to the FIR SED, so by keeping it as a free parameter we allow for the case in which the AGN contributes the majority of the FIR emission.
		We do however restrict the peak of the power law component to fall at shorter wavelengths than the peak of the greybody, which avoids non-physical fits while allowing the possibility that either component dominates the spectrum.
		Nevertheless, our resulting best-fitting average SEDs are consistent with a single warm component plus a significant greybody contribution from cold dust associated with star formation, with all of our fits favouring a significant cold component in the submillimetre spectrum.

		We assume that the cold dust component may be attributed to heating from star-formation processes only, and that the warm dust component is likely have a significant contribution from the AGN.
		Under this assumption, we estimate SFRs using the cold dust component only (see section~\ref{ssub:LAGN}.
		This method neglects any warm dust contribution from star formation, and as such we may underestimate the SFRs.
		As a test, we calculate SFRs using both the warm and cold components, for the case that all FIR emission is due to star formation, which results in SFRs higher than our reported values by a factor of 2 at most.
		Excluding the warm component from the SFR calculation reduces the likelihood of overestimating SFRs in the case of a significant AGN contribution. 

\begin{figure*}
			\subfloat{
				\includegraphics[width=0.25\textwidth]{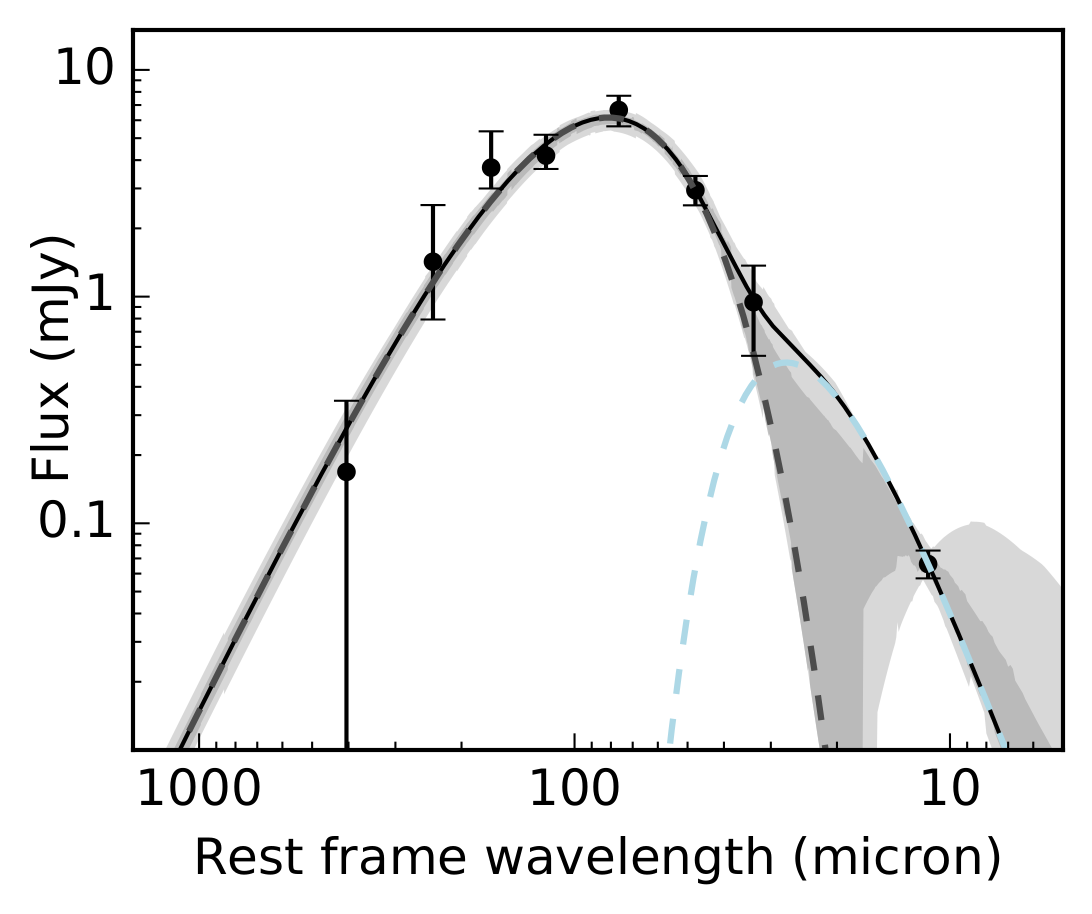}
			}
			\subfloat{
				\includegraphics[width=0.25\textwidth]{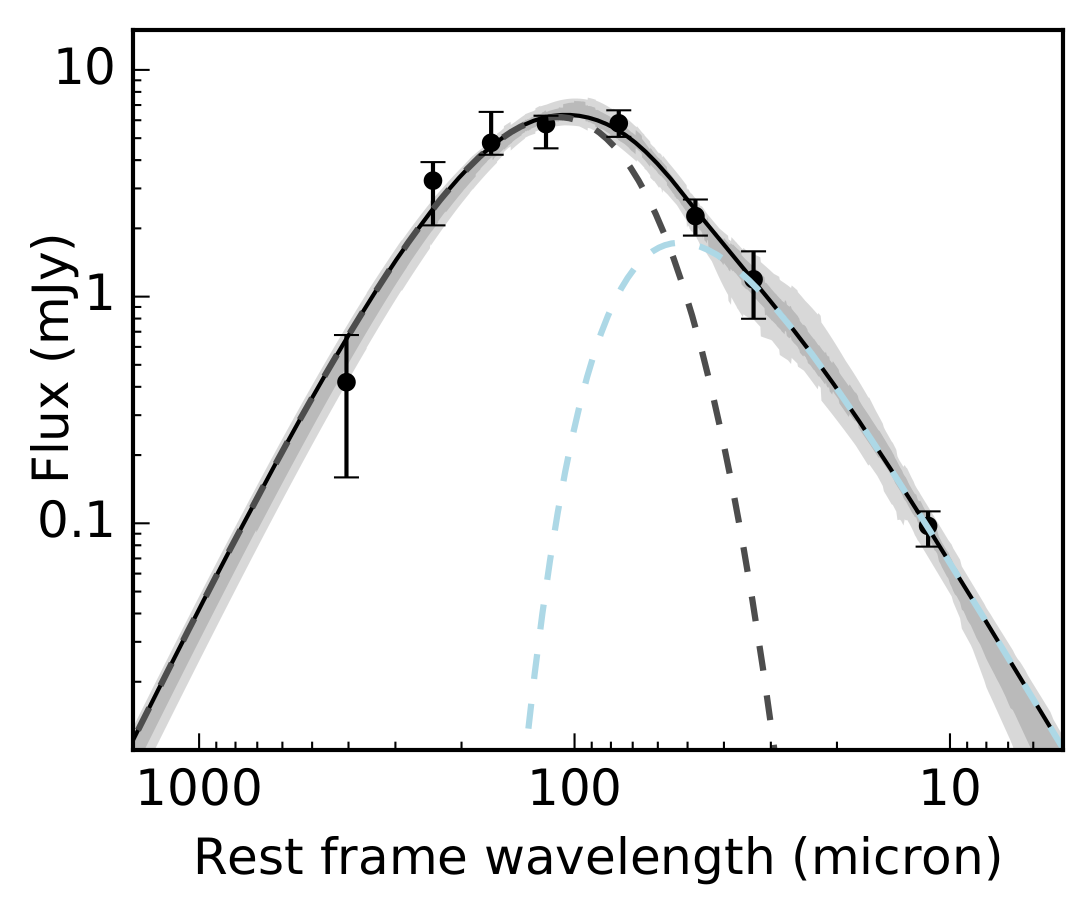}
			}
			\subfloat{
				\includegraphics[width=0.25\textwidth]{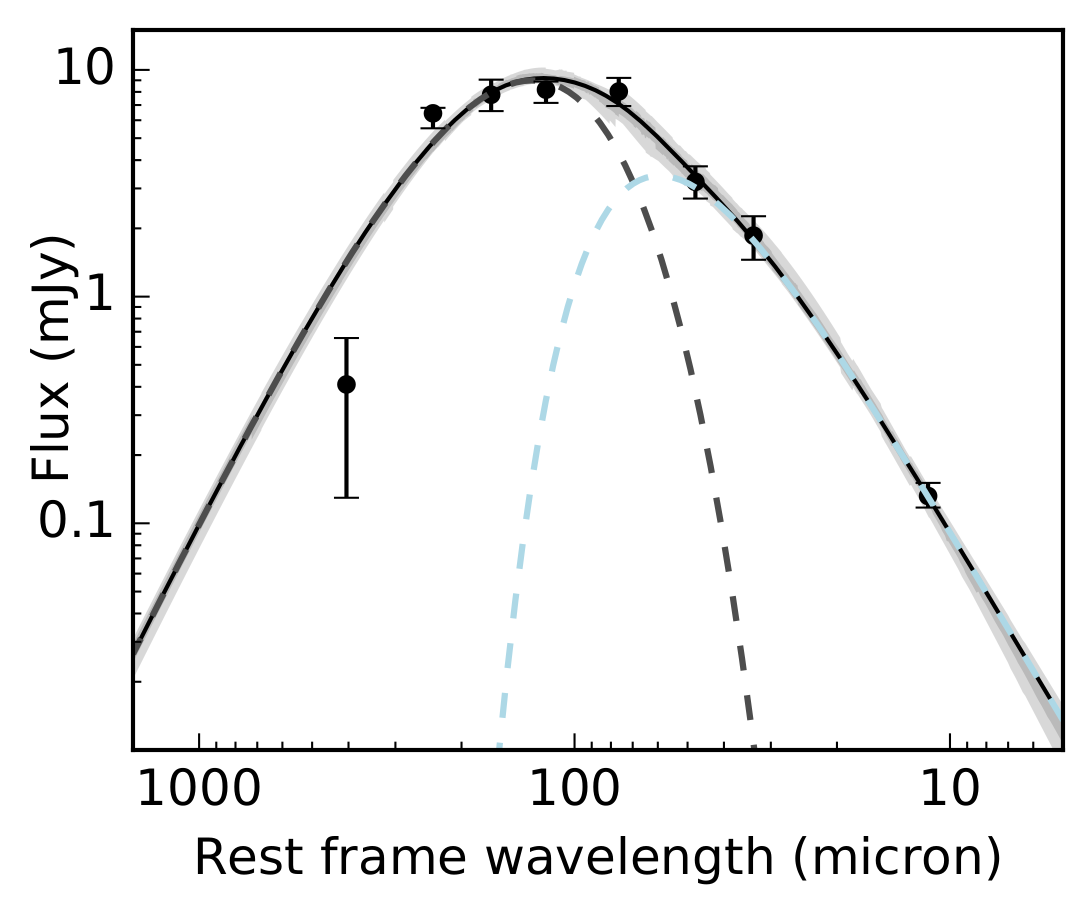}
			}
			\subfloat{
				\includegraphics[width=0.25\textwidth]{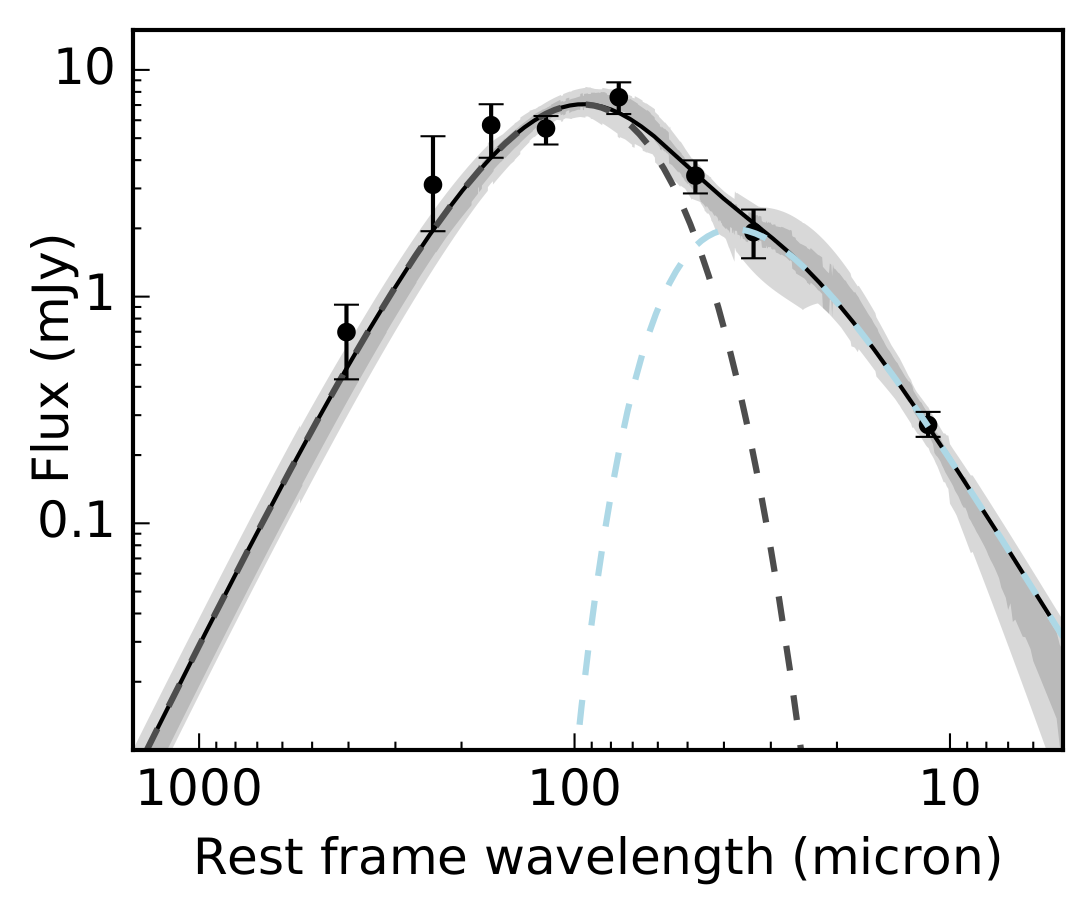}
			}
			\hspace{0mm}
			\subfloat{
				\includegraphics[width=0.25\textwidth]{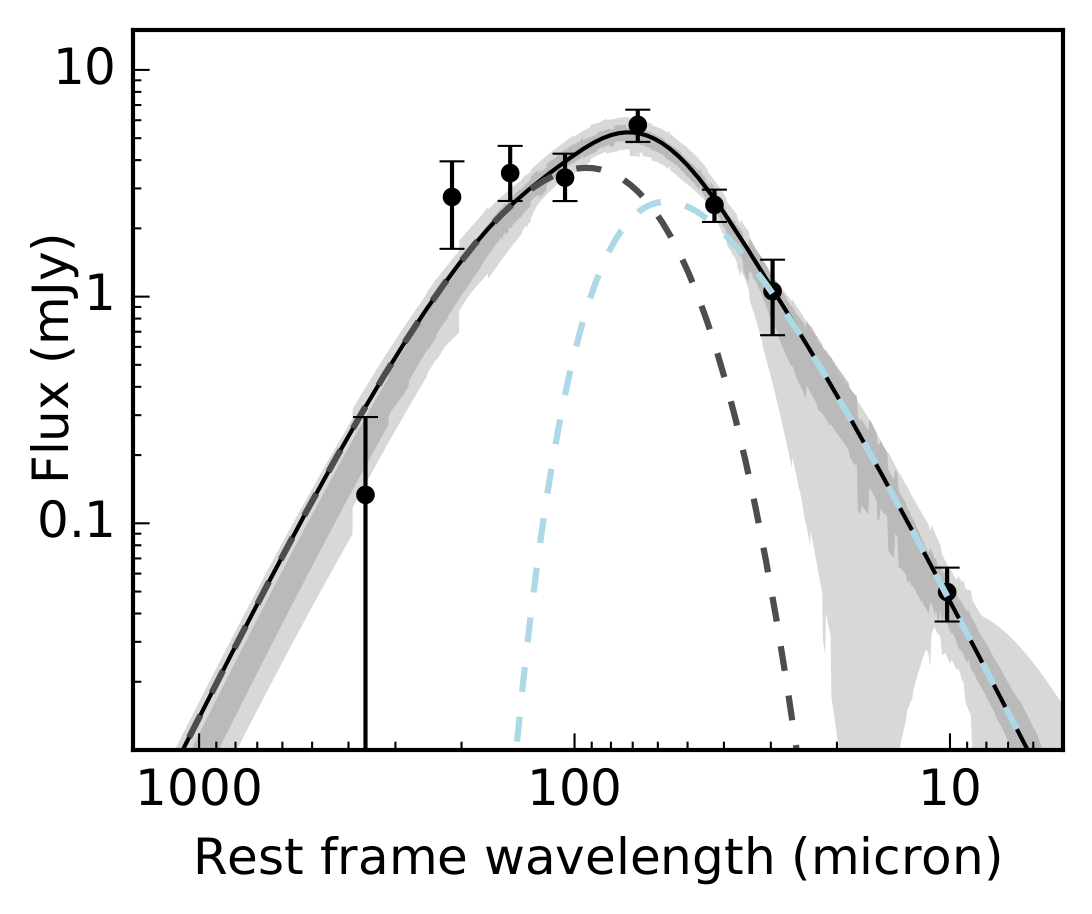}
			}
			\subfloat{
				\includegraphics[width=0.25\textwidth]{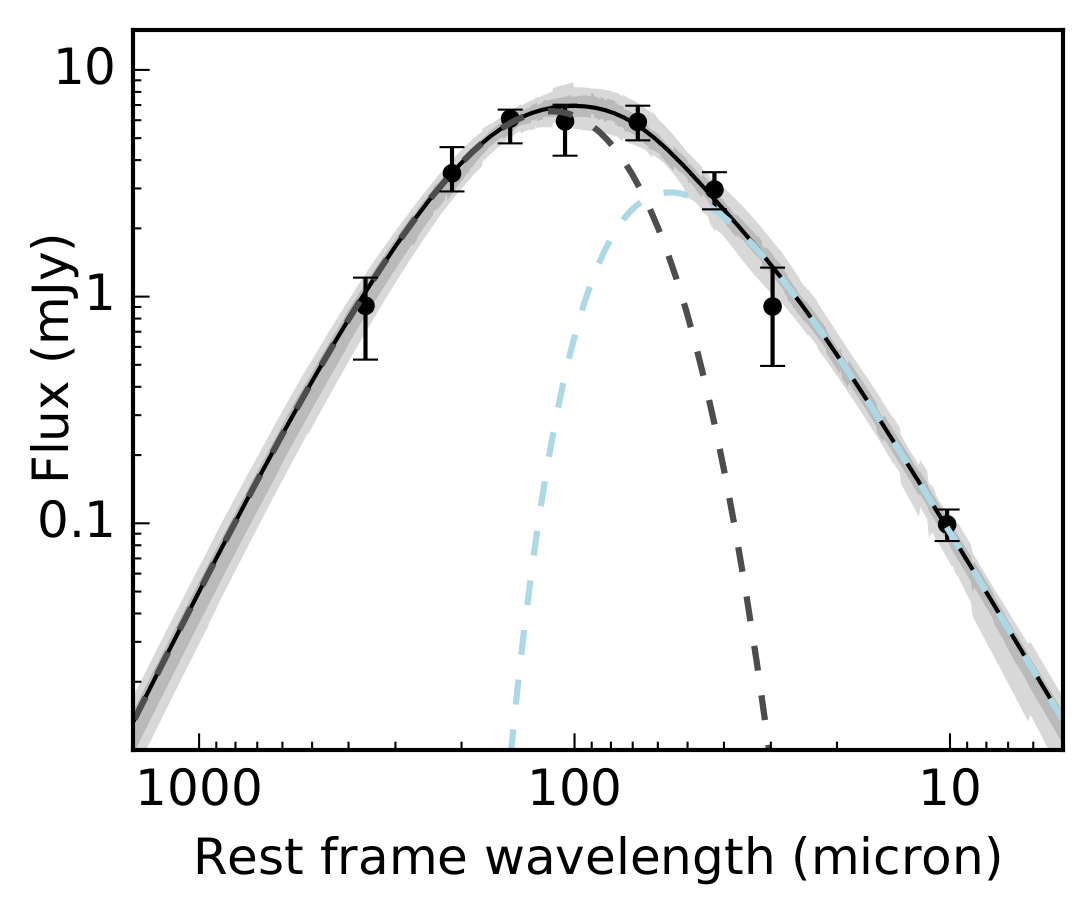}
			}
			\subfloat{
				\includegraphics[width=0.25\textwidth]{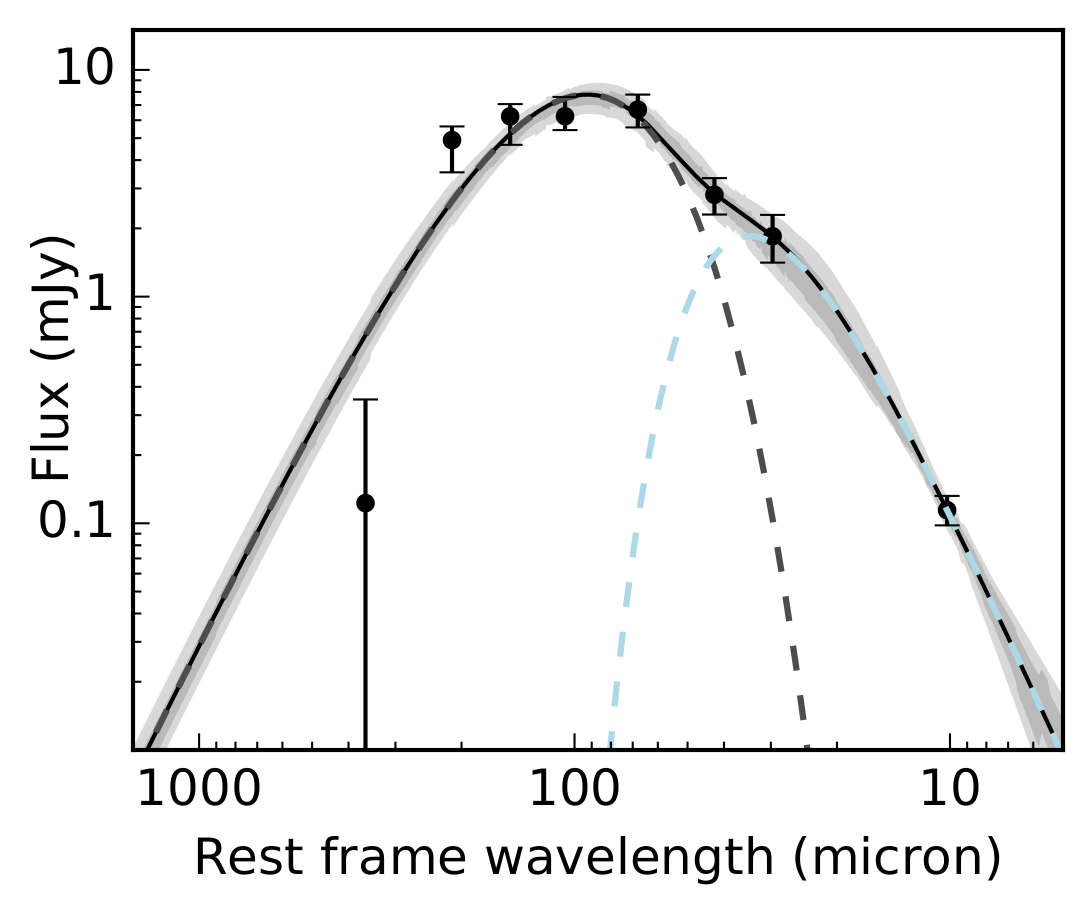}
			}
			\subfloat{
				\includegraphics[width=0.25\textwidth]{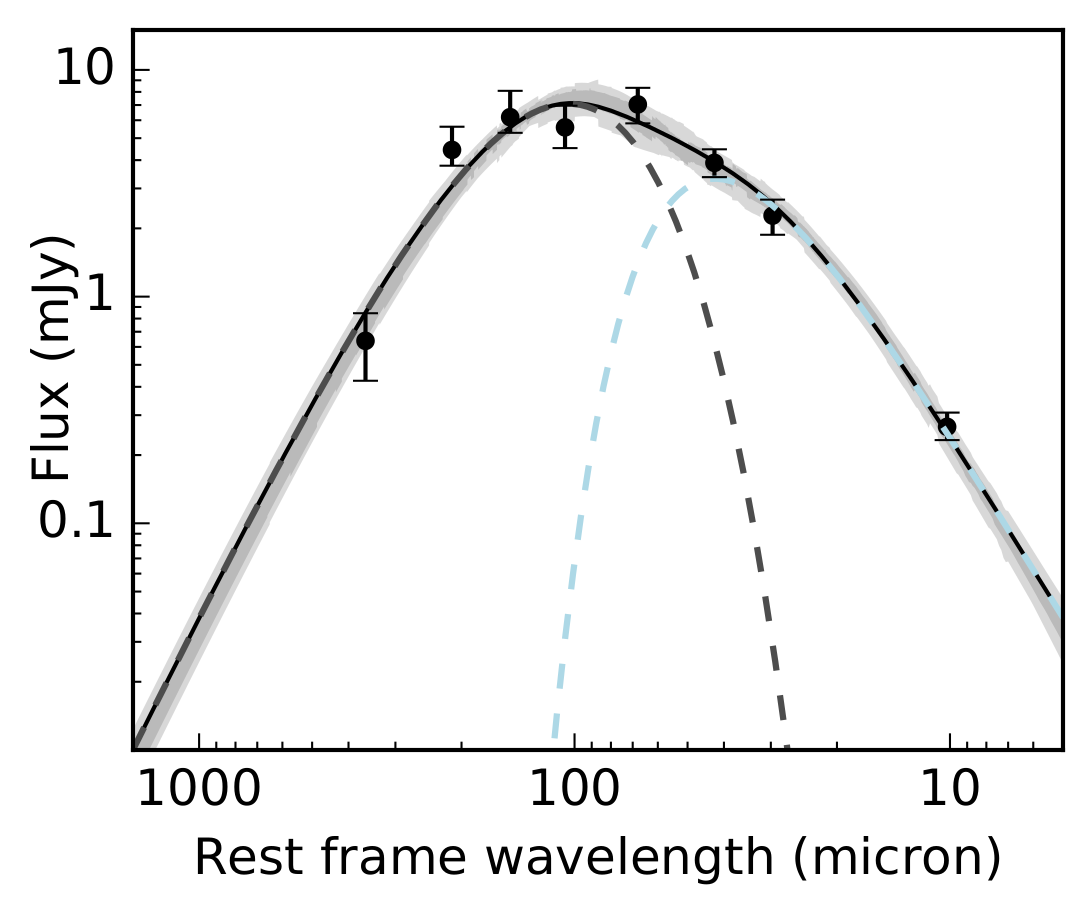}
			}
			\hspace{0mm}
			\subfloat{
				\includegraphics[width=0.25\textwidth]{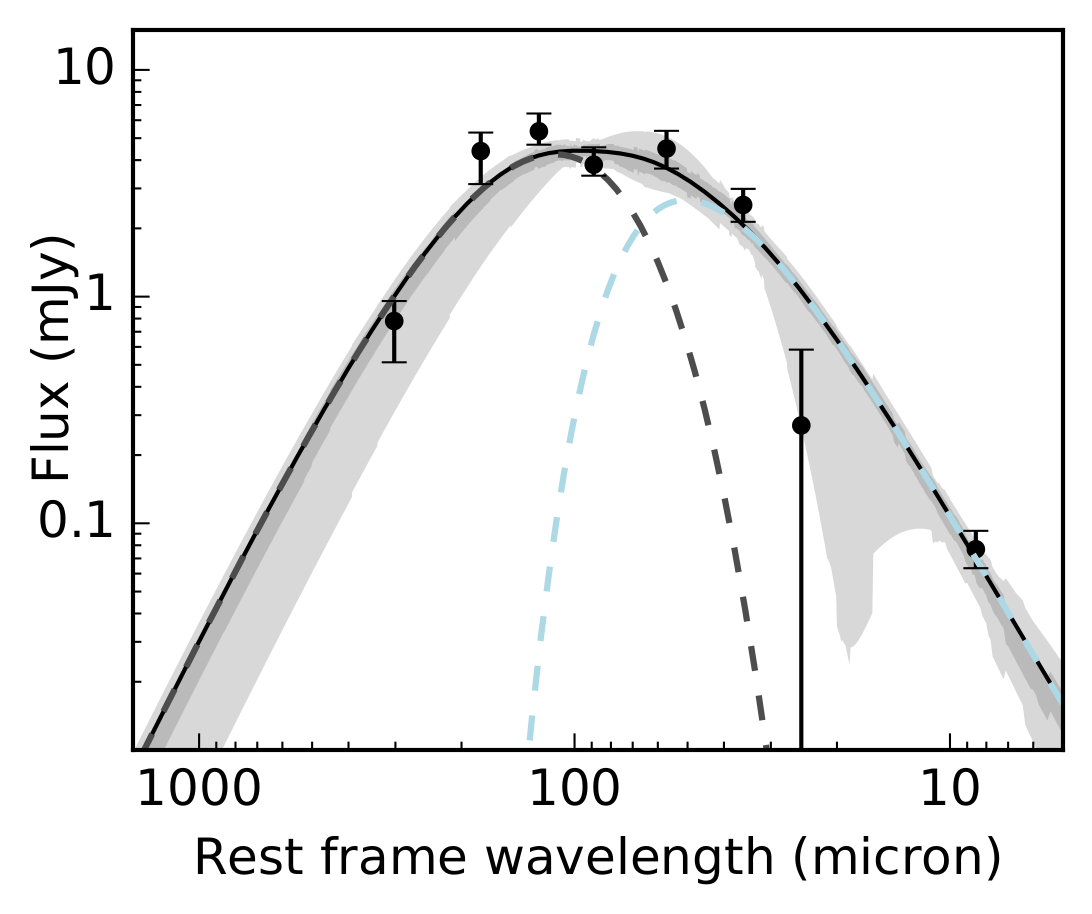}
			}
			\subfloat{
				\includegraphics[width=0.25\textwidth]{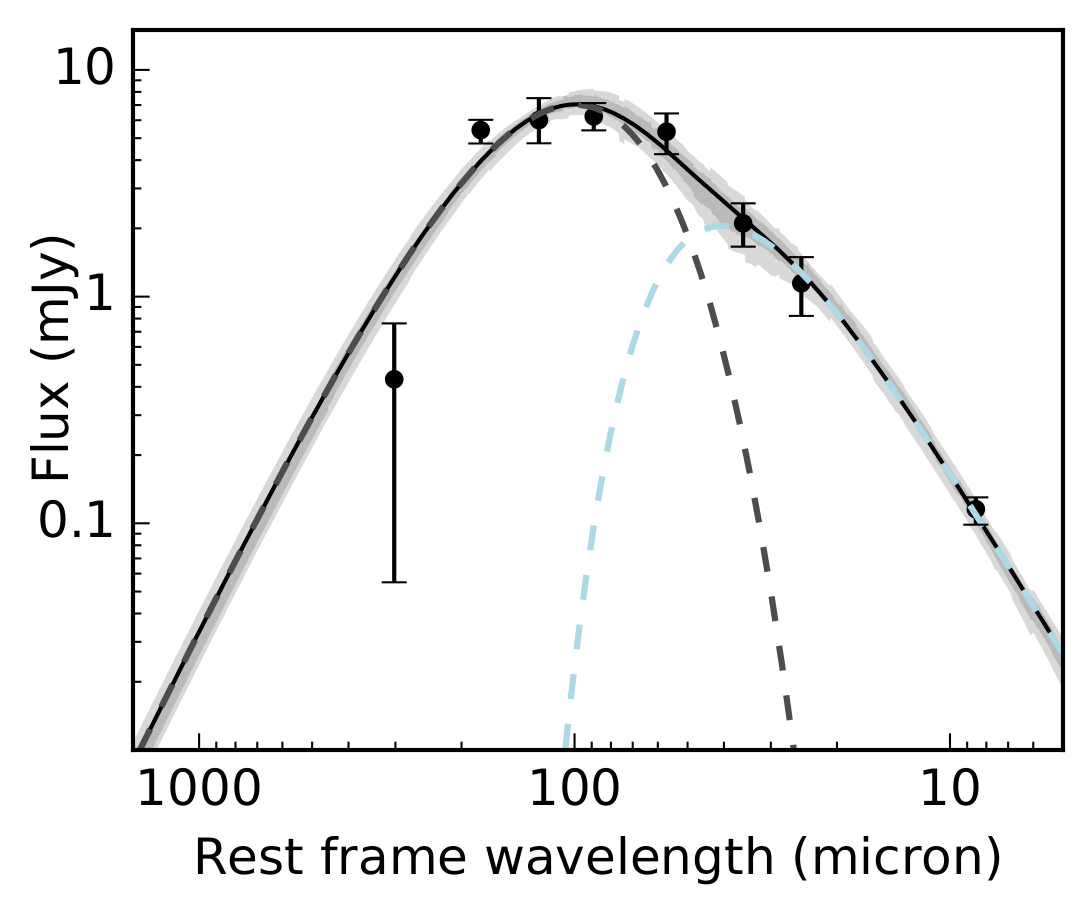}
			}
			\subfloat{
				\includegraphics[width=0.25\textwidth]{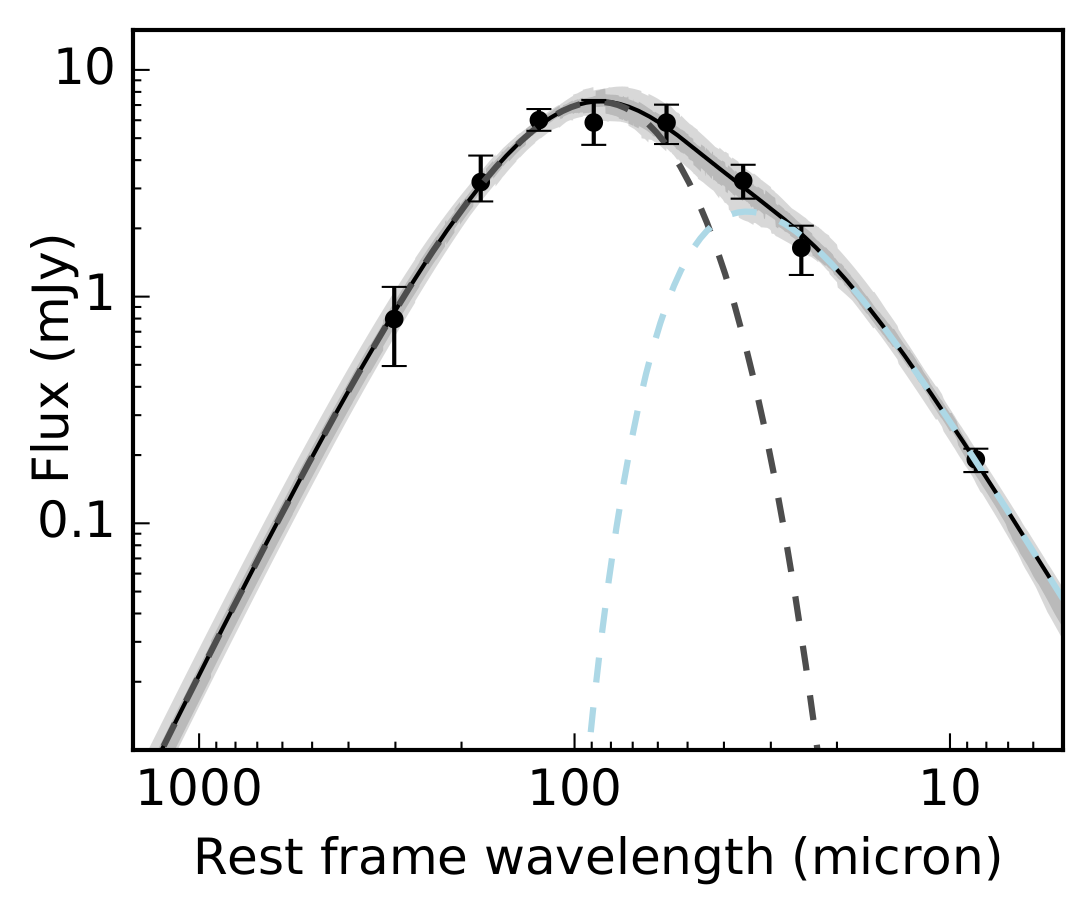}
			}
			\subfloat{
				\includegraphics[width=0.25\textwidth]{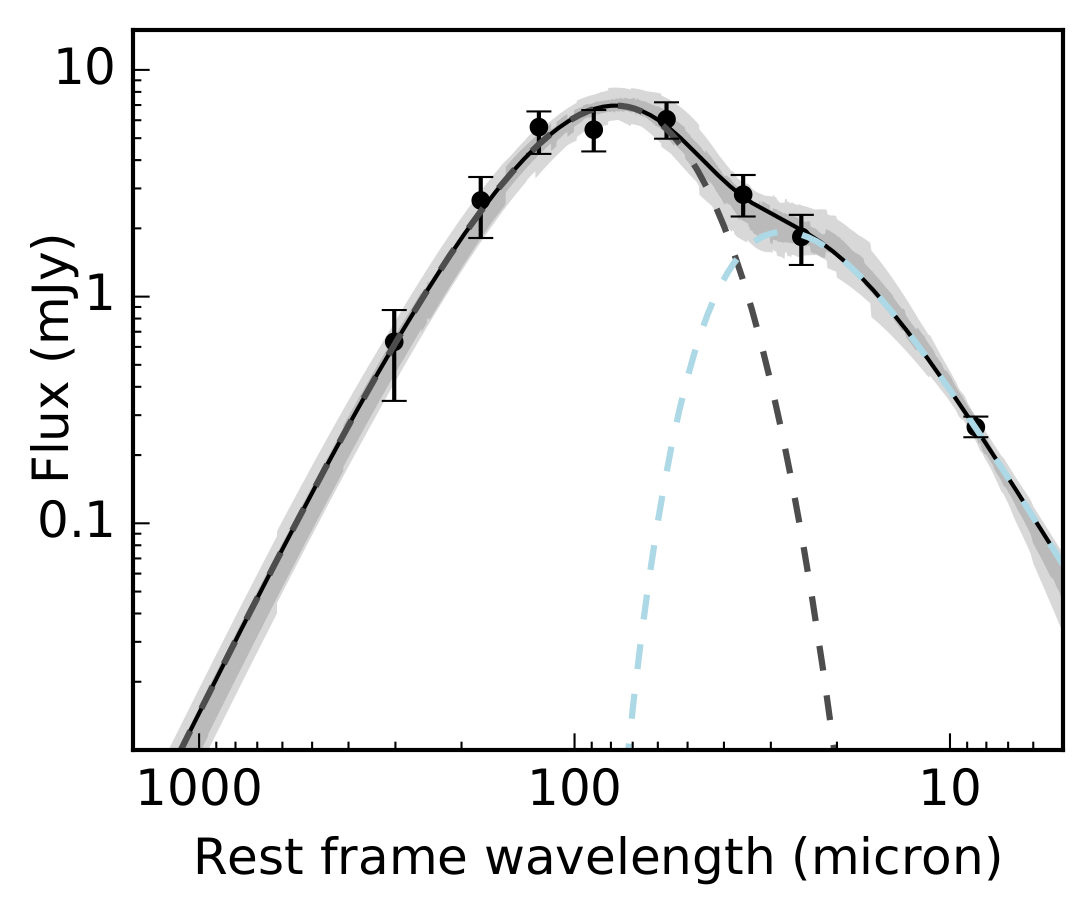}
			}
			\hspace{0mm}
			\subfloat{
				\includegraphics[width=0.25\textwidth]{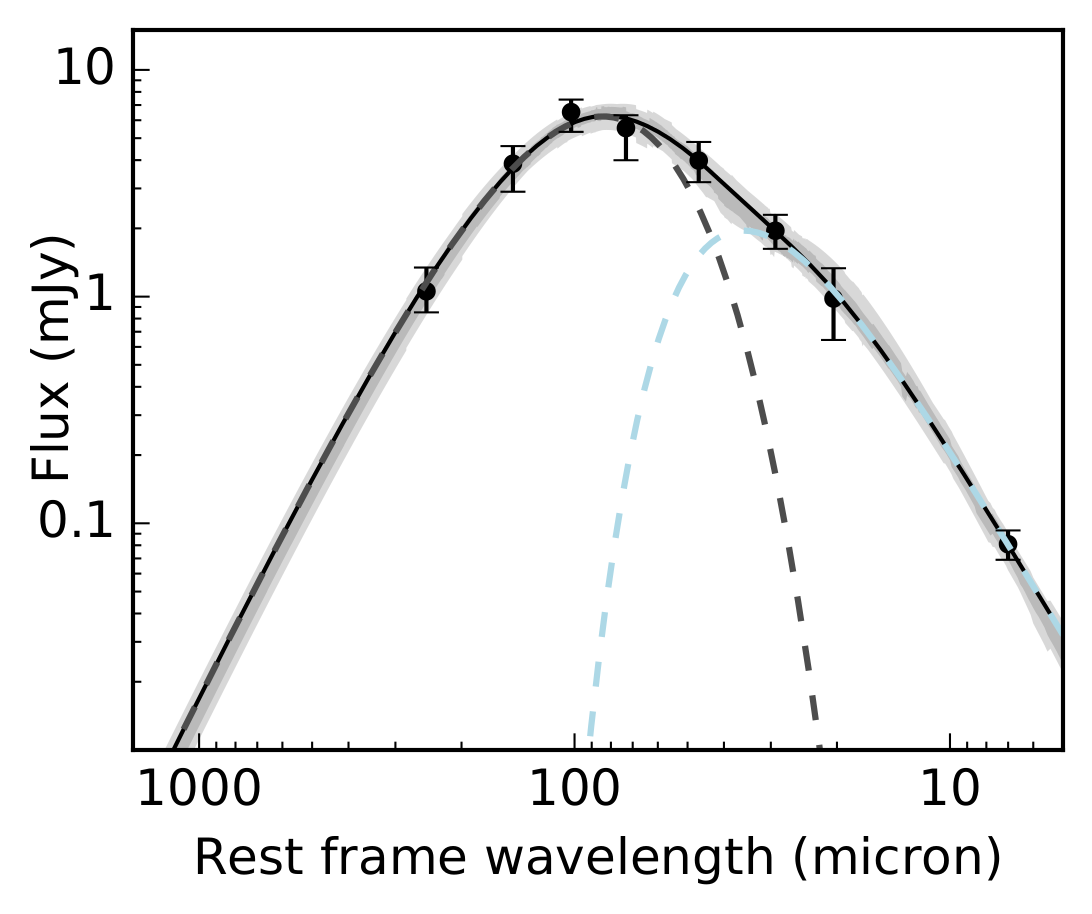}
			}
			\subfloat{
				\includegraphics[width=0.25\textwidth]{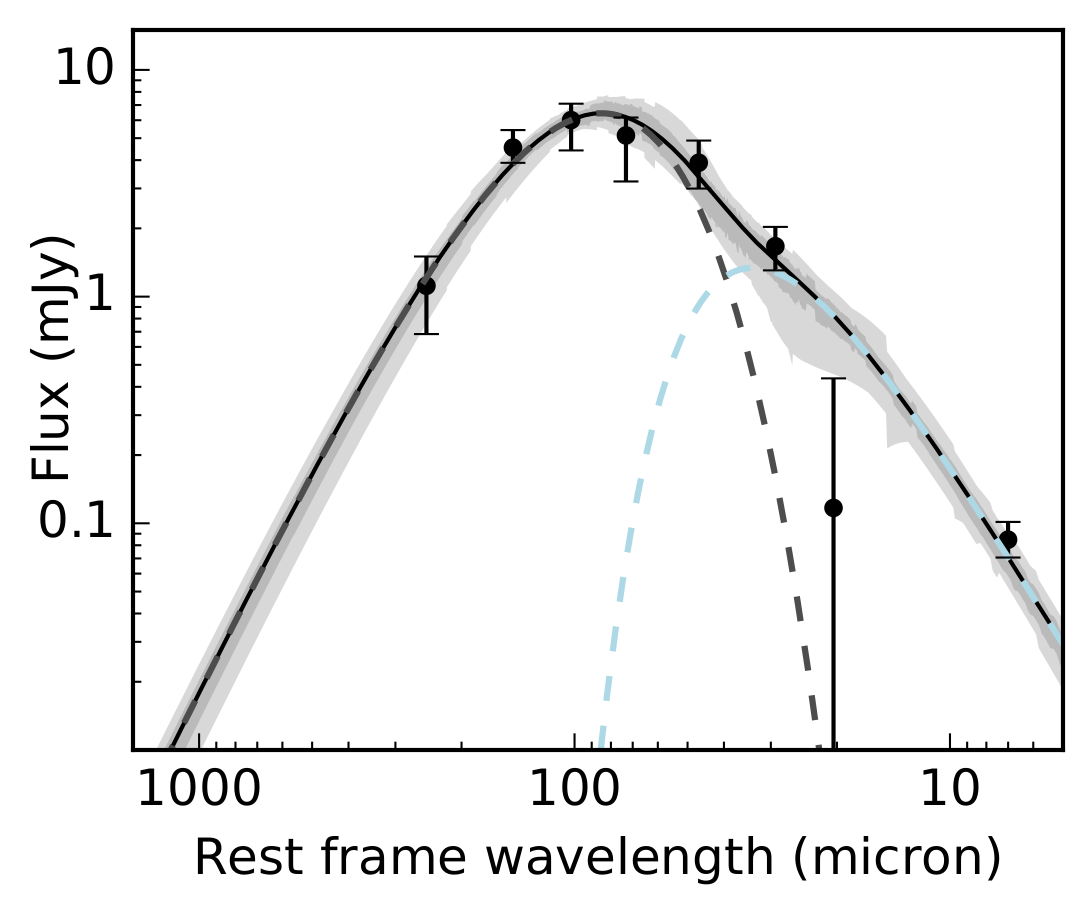}
			}
			\subfloat{
				\includegraphics[width=0.25\textwidth]{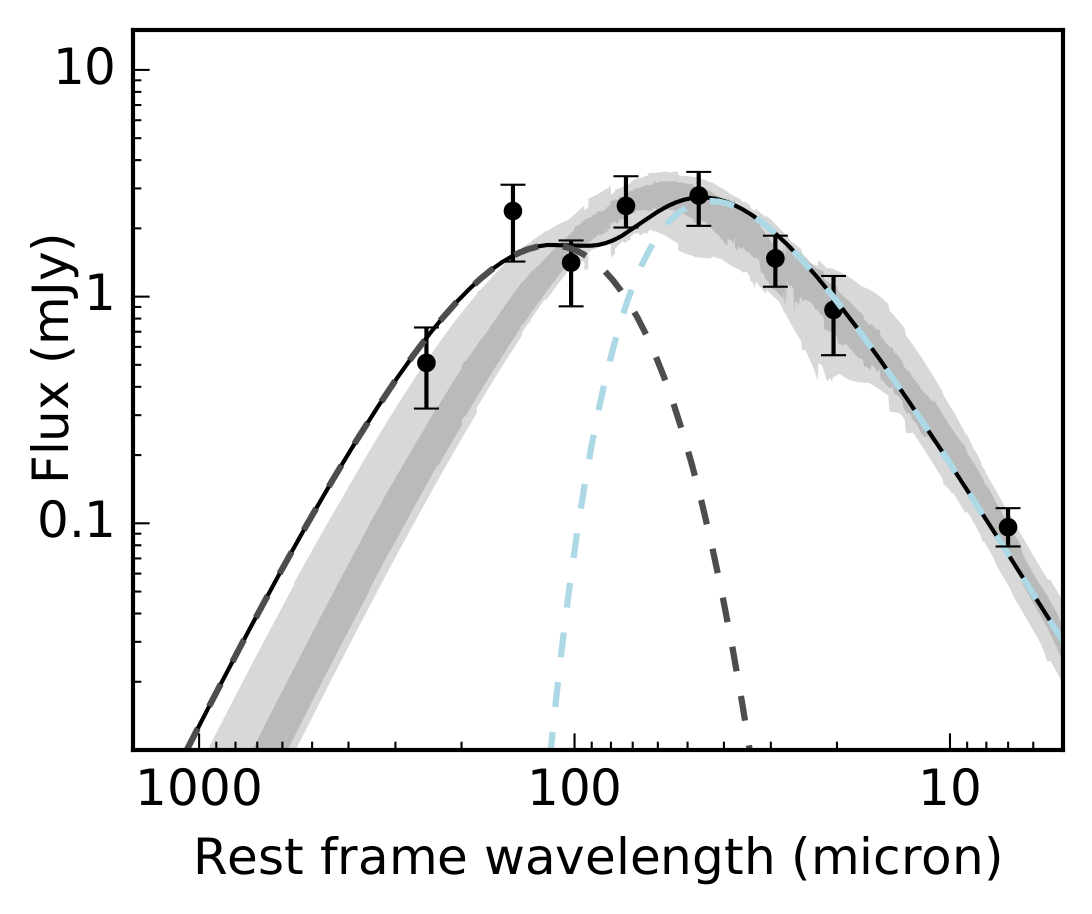}
			}
			\subfloat{
				\includegraphics[width=0.25\textwidth]{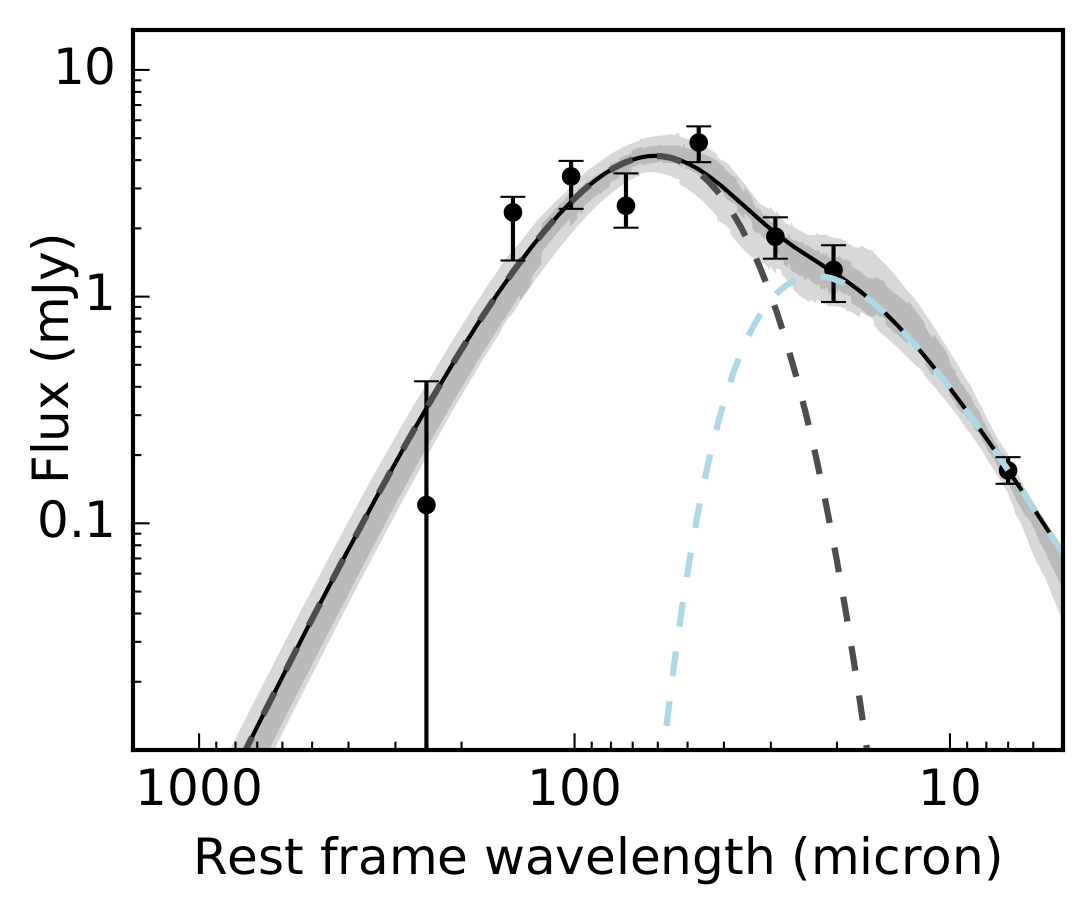}
			}

			\caption{SED fits for each of the 16 `average' sources. Median stacked fluxes are shown as black points, with associated 1$\sigma$ errors determined by a bootstrapping analysis (see section~\ref{ssub:stacking}). The black solid line shows the maximum likelihood realisation, with the grey dashed line showing the cold dust greybody component and the blue dashed line showing the warm dust power law component. The shaded grey regions show the 68 and 95 per cent credible intervals, calculated from our Bayesian analysis.}
			\label{fig:SED_fit}
		\end{figure*}

		We follow an MCMC fitting procedure using the \textsc{emcee} Python code \citep{foreman-mackey_emcee:_2013}.
		This allows a fully Bayesian approach, giving a robust characterisation of the uncertainties of the fit.
		Walkers are distributed with a truncated normal distribution across the prior parameter space, with priors set to exclude non-physical solutions.
		We state the modes of the posterior probability distributions of each parameter in Table~\ref{tab:params}.
		Uncertainties on these values are taken as the Highest Posterior Density (HPD) 68 per cent credible interval.
		The uncertainties on the fit are determined by taking the HPD 68 per cent credible interval at each wavelength.
		Fig.~\ref{fig:SED_fit} shows fitted SEDs for each of the 16 `average' sources, with the median stacked flux data points in black.
		The light blue and grey dashed lines show the warm and cold dust contributions to the fit respectively, with the solid black line showing the maximum likelihood solution.
		The grey shaded areas show the 68 per cent and 95 per cent credible intervals from the HPD at each wavelength.

		\renewcommand{\arraystretch}{1.3}
		\begin{table*}
			\centering
			\caption{Table of fit parameters, showing the modes of the posterior probability distributions for each parameter, for each of the 16 `average' sources fitted, as well as the median redshift $z_{m}$ and X-ray luminosity $L_{\mathrm{X}}$ for each bin as in Table~\ref{tab:bins}. $\mathrm{log}\: N_{\mathrm{gb}}$ is the normalization of the greybody component, $T$ the greybody temperature, $\mathrm{log}\: N_{\mathrm{pl}}$ the normalization of the power law component, and $\alpha $ and $\lambda_{\mathrm{c}}$	are the slope and cutoff wavelength of the power law component respectively. Errors stated are the 68 per cent credible intervals calculated from the HPDs.}
			\label{tab:params}
			\begin{tabular}{ccccccc}
				\hline
				$z_{m}$ & $L_{\mathrm{X}}$ ($10^{43}$ erg\: s$^{-1}$ ) &  $\mathrm{log} N_{\mathrm{gb}}$	& $T $(K) 		& $\mathrm{log} N_{\mathrm{pl}}$	 & $\alpha $	& $\lambda_{\mathrm{c}}$ (\mcm)				 \\
				\hline
				1.10  & $ 1.12 $ & -130.89 $_{-0.20} ^{+0.17}$ & 40.24 $_{-1.60} ^{+1.41}$ & 24.37 $_{-2.97} ^{+13.58}$ & 2.97 $_{-0.25} ^{+1.21}$ & -10.92 $_{-0.83} ^{+0.30}$ \\
				1.10  & $ 2.55 $ & -129.77 $_{-0.24} ^{+0.28}$ & 31.60 $_{-2.07} ^{+2.07}$ & 22.78 $_{-0.79} ^{+7.65 }$ & 2.81 $_{-0.08} ^{+0.66}$ & -10.33 $_{-0.28} ^{+0.28}$ \\
				1.14  & $ 5.95 $ & -128.39 $_{-0.16} ^{+0.13}$ & 25.13 $_{-0.99} ^{+1.02}$ & 22.91 $_{-0.52} ^{+2.03 }$ & 2.80 $_{-0.05} ^{+0.18}$ & -9.94  $_{-0.12} ^{+.09 }$ \\
				1.09  & $ 20.7 $ & -130.27 $_{-0.24} ^{+0.27}$ & 37.00 $_{-2.78} ^{+1.61}$ & 22.95 $_{-0.50} ^{+10.16}$ & 2.74 $_{-0.05} ^{+0.88}$ & -10.57 $_{-0.30} ^{+0.16}$ \\
				\hline
				1.32  & $ 1.87 $ & -131.50 $_{-0.30} ^{+0.36}$ & 44.86 $_{-4.17} ^{+2.86}$ & 25.75 $_{-3.31} ^{+10.82}$ & 3.09 $_{-0.30} ^{+0.94}$ & -10.86 $_{-0.25} ^{+0.98}$ \\
				1.38  & $ 3.84 $ & -129.39 $_{-0.35} ^{+0.26}$ & 29.97 $_{-2.95} ^{+2.22}$ & 23.04 $_{-0.50} ^{+2.54 }$ & 2.81 $_{-0.05} ^{+0.22}$ & -10.06 $_{-0.19} ^{+0.14}$ \\
				1.33  & $ 10.5 $ & -130.14 $_{-0.17} ^{+0.31}$ & 35.29 $_{-2.51} ^{+1.77}$ & 23.99 $_{-1.05} ^{+11.14}$ & 2.85 $_{-0.07} ^{+0.98}$ & -10.59 $_{-0.21} ^{+0.30}$ \\
				1.40  & $ 31.7 $ & -129.80 $_{-0.25} ^{+0.26}$ & 32.57 $_{-2.28} ^{+2.35}$ & 23.23 $_{-0.38} ^{+1.96 }$ & 2.74 $_{-0.04} ^{+0.17}$ & -10.33 $_{-0.11} ^{+0.12}$ \\
				\hline
				1.76  & $ 3.70 $ & -129.82 $_{-0.29} ^{+0.25}$ & 29.21 $_{-1.98} ^{+1.92}$ & 23.40 $_{-0.84} ^{+2.24 }$ & 2.83 $_{-0.08} ^{+0.19}$ & -10.11 $_{-0.10} ^{+0.09}$ \\
				1.76  & $ 7.37 $ & -129.83 $_{-0.25} ^{+0.18}$ & 32.78 $_{-1.58} ^{+1.97}$ & 22.94 $_{-0.45} ^{+2.28 }$ & 2.75 $_{-0.05} ^{+0.19}$ & -10.36 $_{-0.14} ^{+0.13}$ \\
				1.80  & $ 19.3 $ & -130.46 $_{-0.24} ^{+0.22}$ & 38.25 $_{-2.73} ^{+2.06}$ & 23.46 $_{-0.41} ^{+2.50 }$ & 2.74 $_{-0.04} ^{+0.21}$ & -10.50 $_{-0.11} ^{+0.13}$ \\
				1.88  & $ 50.4 $ & -131.03 $_{-0.26} ^{+0.25}$ & 43.24 $_{-3.21} ^{+1.99}$ & 23.47 $_{-0.42} ^{+6.29 }$ & 2.71 $_{-0.04} ^{+0.53}$ & -10.73 $_{-0.20} ^{+0.11}$ \\
				\hline
				2.47  & $ 7.46 $ & -130.74 $_{-0.23} ^{+0.17}$ & 39.14 $_{-1.97} ^{+2.09}$ & 23.66 $_{-0.60} ^{+3.45 }$ & 2.79 $_{-0.05} ^{+0.29}$ & -10.51 $_{-0.14} ^{+0.14}$ \\
				2.53  & $ 15.6 $ & -130.71 $_{-0.27} ^{+0.35}$ & 38.25 $_{-2.79} ^{+3.37}$ & 23.08 $_{-0.69} ^{+1.69 }$ & 2.74 $_{-0.06} ^{+0.15}$ & -10.53 $_{-0.22} ^{+0.15}$ \\
				2.39  & $ 30.0 $ & -133.23 $_{-0.47} ^{+0.64}$ & 57.03 $_{-9.04} ^{+7.03}$ & 23.69 $_{-1.07} ^{+8.68 }$ & 2.77 $_{-0.09} ^{+0.70}$ & -10.91 $_{-0.44} ^{+0.29}$ \\
				2.45  & $ 110  $ & -132.86 $_{-0.24} ^{+0.37}$ & 57.72 $_{-5.31} ^{+3.08}$ & 23.79 $_{-0.46} ^{+11.44}$ & 2.73 $_{-0.04} ^{+0.95}$ & -11.10 $_{-0.24} ^{+0.21}$ \\
				\hline
			\end{tabular}
		\end{table*}
		\renewcommand{\arraystretch}{1}

	\subsection{Star-formation rates with AGN luminosity}
	\label{ssub:LAGN}

		The FIR luminosity due to SF is calculated by integrating under the greybody curve from 8 to 1000 \mcm.
		This is done for each realisation of the fit for each of the 16 `average' sources, taking the mode of the resulting distribution and using the 68 per cent credible interval of the HPD as the uncertainty on the FIR luminosity.

		SFRs are calculated from this FIR luminosity using the \citet{kennicutt_star_1998} relation:
		\begin{equation}
			{\mathrm{SFR}} (M_{\odot}{\mathrm{yr}}^{-1}) = 4.5 \times 10^{-44} L_{\mathrm{FIR}} ({\mathrm{erg\:s}}^{-1})
		\end{equation}		

\begin{figure*}
			\includegraphics[width=0.8\textwidth]{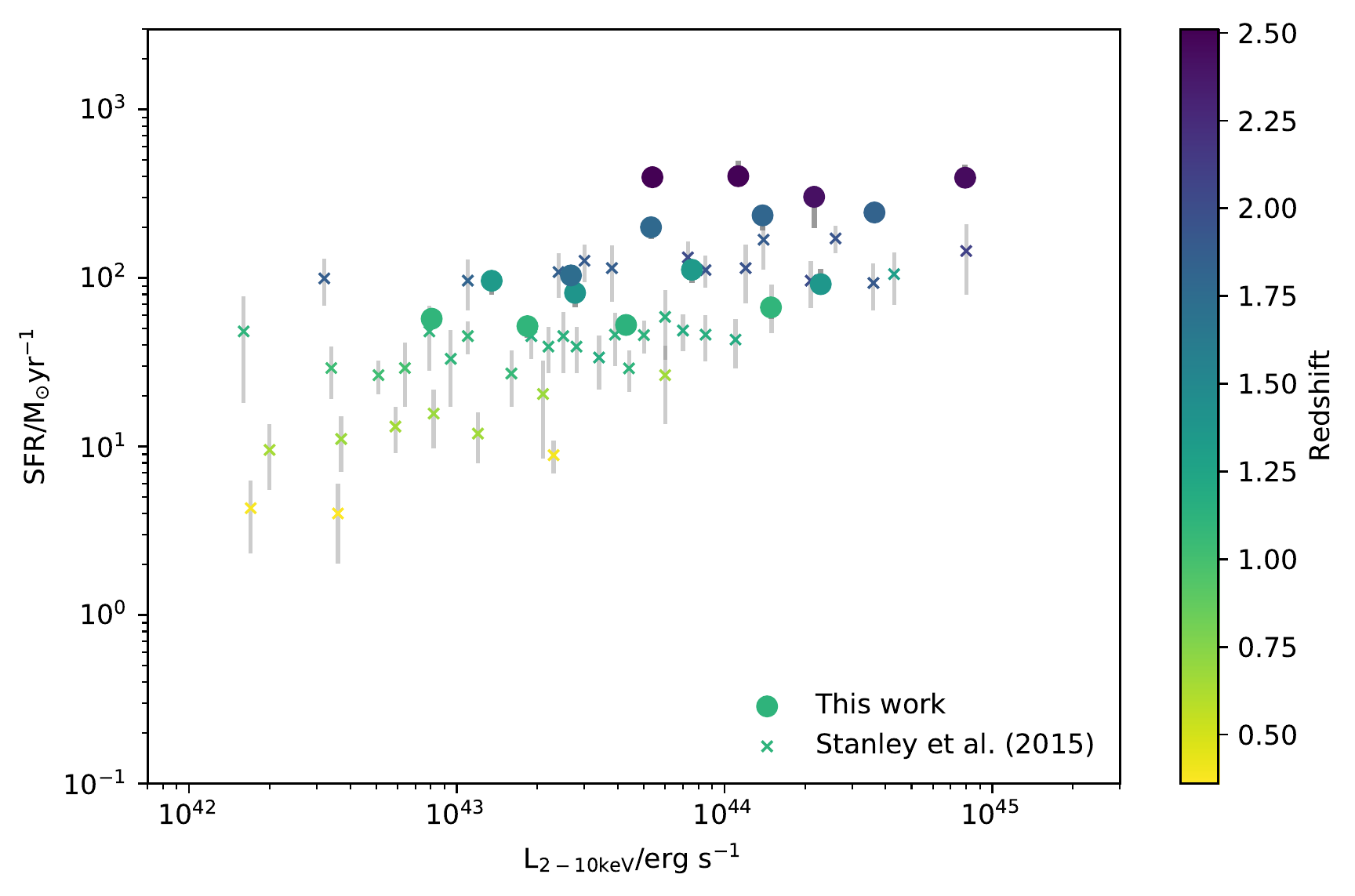}
			\caption{SFR versus X-ray luminosity, for each of the 16 median `average' sources. Circles show this work, with 1$\sigma$ errors; where not visible, errors are within the size of the points. Crosses are from \citet{stanley_remarkably_2015}. Colours indicate redshift; the \citet{stanley_remarkably_2015} sample covers a redshift range $0.2 < z < 2.5$. There is no evidence for a decrease in SFR with increasing AGN power, and within each redshift bin there is no significant positive or negative trend of SFR with $L_{\mathrm{X}}$.}
			\label{fig:Lir_Lagn}
		\end{figure*}

		Fig.~\ref{fig:Lir_Lagn} shows the resulting SFRs calculated for each of the 16 average sources for each redshift and $L_{\mathrm{X}}$ bin.

		For comparison, we also plot the results of \citet{stanley_remarkably_2015}.
		Following \citet{xue_2_2016}, we convert our $L_{0.5-7~\rm keV}$ luminosities to equivalent $L_{2-10~\rm keV}$ luminosities for the purposes of comparison, assuming a photon index $\Gamma = 1.8$, such that:
		\begin{equation}
			L_{2-10~\rm keV} = 0.721 \times L_{0.5-7~\rm keV}
		\end{equation}

		Our results are in good agreement with the study by \citet{stanley_remarkably_2015}, which follows a similar approach with several notable differences. 
		We use analytical models in our SED fitting procedure rather than template fitting.
		As discussed in section~\ref{ssub:SED}, this removes the issue of biases regarding the choice of templates used in the fit, however may underestimate SFRs if there is a significant warm dust component from SF.
		We also make use of newer, deeper X-ray surveys \citep[e.g.][]{xue_2_2016, civano_chandra_2016} and as such have a larger sample of sources, so are able to use only spectroscopic redshifts thus more accurately determined X-ray luminosities.
		Our study extends to slightly higher redshift, spanning an additional $\sim$ 0.5 Gyr in cosmic time, and expanding the parameter space in which this relationship has been investigated.
		Additionally, using the S2CLS data provides additional information about the Rayleigh-Jeans tail of the dust spectrum, allowing us to closely constrain the fit at long wavelengths.
		We also follow a median rather than mean stacking approach, which avoids the issue of the few brightest sources in the stack biasing the stacked flux to a higher value \citep[e.g.][]{barger_host_2015}.
		Despite these differences in method, our results do agree with those of \citet{stanley_remarkably_2015}, displaying the same flat trend of SFR with $L_{\mathrm{X}}$.

		To investigate the systematics of our method, we construct an SED template using \citet{siebenmorgen_2007} starburst and \citet{siebenmorgen_2015} AGN torus templates, and extract fluxes at the observed wavelengths.
		Applying our fitting mechanism to these simulated datapoints, we find that we consistently underestimate the template SFRs by a factor of $\sim 30$ per cent.
		As described in section~\ref{ssub:SED}, this is to be expected as we do not account for any contribution to the MIR spectrum from SF.
		However, as we are largely interested in the trend of SFR with $L_{\mathrm{X}}$ rather than absolute SFRs, this systematic discrepancy does not affect our overall conclusion.
		We also investigate whether varying $\beta$ has a significant impact on our fitted SFRs; across the range $1 < \beta < 2$ we find the resulting SFR varies by a factor of $\sim 10$ per cent and similarly does not affect our conclusion.

		As in section~\ref{ssub:850_stack}, we find no decrease in SFR with increasing $L_{\mathrm{X}}$.
		Across the redshift range of the whole sample, there is a mild positive slope with SFR increasing with $L_{\mathrm{X}}$.
		In previous work \citep[e.g.][]{stanley_remarkably_2015, stanley_mean_2017}, this positive slope has been attributed to the evolution of SFR with stellar mass (the `main sequence' of star-forming galaxies).
		We do not have reliable stellar masses for the galaxies in our sample, but referring to their work, this explains well the trend that is observed.
		In \citet{stanley_mean_2017}, stellar mass is inferred from black hole mass $M_{\bullet}$.
		Selecting sources by their X-ray flux results in the selection effect that the highest $L_{\mathrm{X}}$ sources in our sample are those objects that have both the highest Eddington ratios, and the highest $M_{\bullet}$, thus higher stellar masses, assuming the stellar mass -- $M_{\bullet}$ relation.
		Galaxies of higher $L_{\mathrm{X}}$ are therefore typically those with higher $M_{\bullet}$ and inferred stellar mass, which corresponds to higher SFR.
		Additionally, with increasing redshift bins we have both the selection effect that we select a greater number of more luminous sources when probing a larger volume, and the effect of the redshift evolution of the SF main sequence \citep[e.g.][]{noeske_star_2007}.
		Thus we can draw the conclusion, following \citet{stanley_mean_2017}, that the AGN in our sample reside in normal star-forming galaxies, and the trend of SFR -- $L_{\mathrm{X}}$ can be attributed to the star forming main sequence of galaxies.
		Within each redshift bin, the SFR is consistent with a flat trend with $L_{\mathrm{X}}$, within errors.
		The only exception is the $1.53 < z < 2.05$ bin in which there is a slight positive trend.
		
		Notably, we do not see any negative correlation between SFR and $L_{\mathrm{X}}$ that we might interpret as evidence that the most powerful AGN quench star formation in their host galaxies.
		This work adds to a number of recent studies that have markedly not observed this anticorrelation \citep[e.g.][]{stanley_mean_2017, suh_type_2017, scholtz_identifying_2018}.

		There are a number of explanations for this lack of signature.
		It may be due to the time-scales of AGN activity relative to star formation \citep{hickox_black_2014, harrison_impact_2017}: interpreting this relationship as evidence of negative AGN feedback or lack thereof requires the assumption that instantaneous X-ray luminosity is a reliable indicator of AGN activity.
		However, X-ray luminosity can vary by several orders of magnitude over very short time-scales ($10^{3} - 10^{6}$ years) due to the stochastic nature of AGN fueling, relative to much longer time-scales associated with star formation and quenching ($\sim 10^{8}$ years, e.g. \citealp{hickox_black_2014}).
		As such, instantaneous X-ray luminosity does not reliably trace long-term average BHAR, and we do not necessarily expect to see a negative correlation between SFR and $L_{\mathrm{X}}$ as an indication of suppressed star formation.

		One way to investigate the observational effects of AGN feedback-driven quenching is by comparing observations to the results of large-scale cosmological simulations.
		We perform a simple comparison, using the results of the \textsc{Horizon-AGN} simulation \citep{dubois_dancing_2014}.

		\begin{figure*}
			\includegraphics[width=\textwidth]{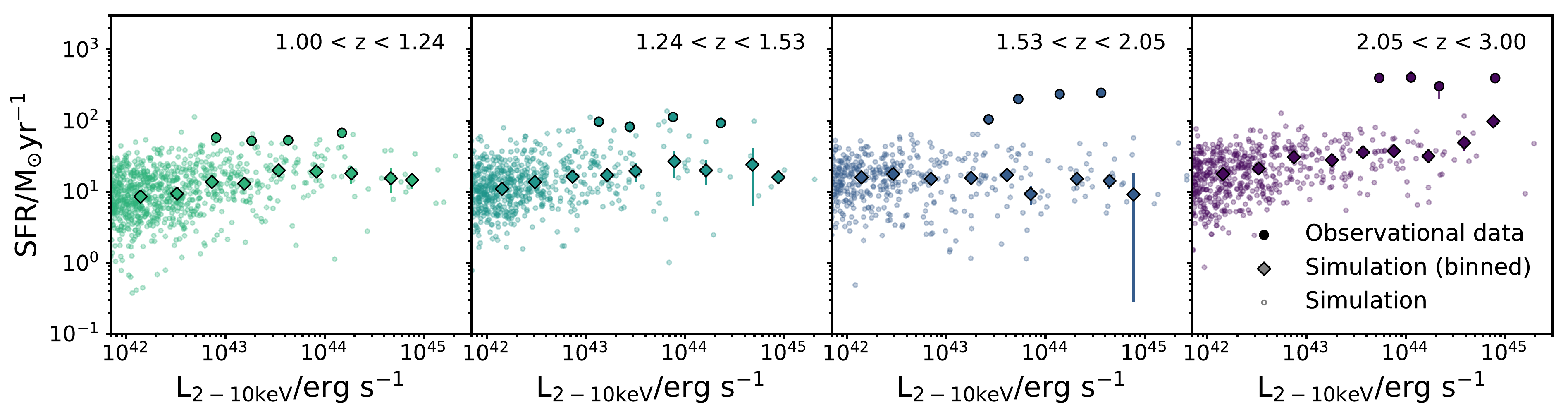}
			\caption{SFR versus X-ray luminosity, for both the observational data (filled circles) and the simulated \textsc{Horizon-AGN} galaxies (dots, with binned values shown with filled diamonds), for each redshift bin. Colours correspond to redshifts in Fig.~\ref{fig:Lir_Lagn}. Both observations and simulations show, within each redshift bin and across this $L_{\mathrm{X}}$ range, a flat trend of SFR with increasing $L_{\mathrm{X}}$.}
			\label{fig:sim_Lir_LX}
		\end{figure*}

		\subsubsection{The \textsc{Horizon-AGN} simulation}
		\label{ssubsub:Horizon-AGN}

			The \textsc{Horizon-AGN} simulation is large-scale hydrodynamical cosmological simulation, tracing the stellar mass growth of galaxies from redshift $z\sim6$.
			It successfully reproduces observable galaxy properties such as the stellar mass and luminosity functions and cosmic star formation history \citep{Kaviraj2017}, as well as BH demographics such as the cosmic evolution of BH mass density and BH -- host galaxy correlations \citep{Volonteri2016, Martin2018}.
			Although the model is described in detail in \citet{dubois_dancing_2014} and \citet{Kaviraj2017}, we briefly outline the treatment of BHs and star formation here as these are relevant to the observable quantities in this study.

			Star formation follows a standard \citet{Kennicutt1998} law with 2 per cent efficiency, when the hydrogen density exceeds $n_{0}=0.1$~H~cm$^{-3}$.
			Stellar feedback is implemented via a subgrid model which includes all processes that contribute to thermal and kinetic feedback on ambient gas \citep[see][]{Kaviraj2017}.

			BHs with initial mass 10$^{5}$ $M_{\odot}$ are seeded as `sink' particles where local gas density exceeds $\rho > \rho_{0}$ , and stellar velocity dispersion exceeds 100~km~s$^{-1}$, where $\rho_{0}=1.67\times10^{-25}$~g~cm$^{-3}$ and corresponds to 0.1~H~cm$^{-3}$.
			BHs grow through both gas accretion, which follows a Bondi-Hoyle-Lyttleton model with accretion capped at the Eddington limit, and through mergers \citep{dubois_dancing_2014,dubois_horizon-agn_2016}.

			Two modes of BH feedback are implemented, based on the accretion efficiency $\chi=\dot{M}_{\mathrm{BH}}/\dot{M}_{\mathrm{Edd}}$.
			At low accretion efficiencies ($\chi<0.01$), a `radio' mode feeds energy into the surrounding medium via bipolar outflows, with jet velocities of $10^4$~km~s$^{-1}$.
			A `quasar' mode takes over at high accretion efficiencies ($\chi<0.01$), modelled as an isotropic thermal energy injection of 1.5 per cent of the accretion energy into the surrounding gas.

		\subsubsection{Bolometric correction}
		\label{ssubsub:bolometric_correction}
			We calculate BH bolometric luminosities assuming a radiative efficiency of 10 per cent \citep{Shakura1973} and use the recommended bolometric corrections for radio-loud and radio-quiet quasars from \citet[][Equation 14 and 15]{Runnoe2012} to convert these bolometric luminosities to $L_{2-10~\rm keV}$.
			For BHs with Eddington ratios where $\chi>0.01$ (quasar-mode) we use the radio-loud correction, while for those with Eddington ratios where $\chi<0.01$ (jet-mode) we use the radio quiet correction.
			Using these corrections, we are able to reproduce the approximate normalization (to within 1 dex, across the full range of $L_{\mathrm{X}}$ considered in this study) and slope of the observed X-ray luminosity function at $z \sim 4$ \citep[e.g.][]{Georgakakis2015}.

		\subsubsection{Comparison with observational result}
		\label{ssubsub:comparison_sim_obs}

			Galaxies in the \textsc{Horizon-AGN} simulated sample are split into bins of 0.5 dex in $L_{\mathrm{X}}$ in the four redshift intervals of the observational sample.
			Fig.~\ref{fig:sim_Lir_LX} shows SFR plotted against $L_{\mathrm{X}}$, as in Fig.~\ref{fig:Lir_Lagn}, for each bin of the simulated dataset (open diamonds) and the observational result (closed circles; colours correspond to redshift bins).
			The simulated results, incorporating the AGN feedback model described in Section~\ref{ssubsub:Horizon-AGN}, also show a flat trend of SFR with increasing $L_{\mathrm{X}}$.
			It should be noted that, while the trends agree, there is around an order of magnitude discrepancy between the normalization of SFR in the simulation compared to the observations.
			This discrepancy is not unusual in semi-analytical models and hydrodynamical simulations \citep[e.g.][]{Lamastra2013, sparre_2015}, and one reason for this may be the star formation prescription.
			While simulations are capable of reproducing averaged properties such as SFR density, the models of star formation that are used may not accurately reproduce instantaneous star formation properties of individual galaxies, specifically the high SFRs associated with galaxies undergoing a starburst phase.
			Star formation in the simulation assumes a star formation efficiency of 2 per cent, an assumption that may not be valid across all redshifts.
			Alternatively, a more complex model of star formation may result in higher peak SFRs.

			Nevertheless, SFR shows no increasing or decreasing trend with $L_{\mathrm{X}}$.
			This is in agreement with \citet{mcalpine_link_2017} and \citet{scholtz_identifying_2018}, who compare the SFR, specific SFR and X-ray luminosity of a sample of AGN with an equivalent simulated sample from the EAGLE suite of hydrodynamical simulations \citep{crain2015}.
			They find that their observational results match closely to that of the simulated sample, whether or not the simulation includes AGN feedback.
			In neither sample do they observe a trend of decreasing SFR with X-ray luminosity that one might expect in the case of AGN quenching star formation, instead also finding a flat relation.
			The AGN feedback model implemented in the \textsc{Horizon-AGN} simulation is somewhat more complex than that of EAGLE, which incorporates only one mode of AGN feedback; we also determine SFRs using a different method (as outlined in Section~\ref{ssub:LAGN}), and our observational dataset is an order of magnitude larger than that of \citet{scholtz_identifying_2018} and extends to higher redshifts than those considered in \citet{mcalpine_link_2017}.
			Despite these differences in approach in both simulation and observation, our result agrees with this previous work.
			This provides another piece of evidence to suggest that, even with AGN feedback as a fundamental mechanism to quench star formation in galaxy growth, we do not expect to see the observational signature of this in the relation betweeen SFR and instantaneous AGN X-ray luminosity.

		\subsubsection{No enhanced star formation at high $L_{\mathrm{X}}$}
		\label{ssubsub:no_enhanced}

			There is also no sign of \textit{enhanced} SFR in the most luminous AGN, as reported in previous work \citep[e.g.][]{lutz_laboca_2010, rosario_mean_2012}.
			These studies find a strong correlation between SFR and AGN luminosity only in the most luminous AGN (generally, $L_{\mathrm{AGN}} \gtrsim 10^{44} {\mathrm{erg\: s}}^{-1}$).
			This has been interpreted as evidence of two modes of AGN -- galaxy coevolution, in the two different $L_{\mathrm{AGN}}$ regimes: a secular evolution of SF in the low luminosity regime, and a tight AGN -- SFR coevolution in the high luminosity regime.
			However, studies that find this positive SFR -- $L_{\mathrm{AGN}}$ correlation in high luminosity AGN have only examined low redshift ($z < 1$) samples; here, we do not find any evidence of this trend at $z > 1$.
			In the context of this two-mode model, it is this secular mode of evolution that gives rise to the flat relationship that we observe between SFR and $L_{\mathrm{X}}$ within each redshift bin.

\section{Conclusions}
\label{sec:conclusion}

	We investigate the FIR and submm properties of a sample of $\sim$1000 X-ray selected AGN, in the redshift range $1 < z < 3$.
	Data from S2CLS, \textit{Herschel} and \textit{Spitzer} across 8 wavelengths are used to constrain the FIR spectrum. 
	The fraction of sources detected in the submm across four $L_{\mathrm{X}}$ bins is investigated, and a simple stacking approach is used to analyse the submm properties, in which we find no significant increase or decrease of submm flux with $L_{\mathrm{X}}$.
	We stack sources in bins of redshift and X-ray luminosity, to create 16 `average' sources across the parameter space.
	Using a Bayesian MCMC method, we fit SEDs to each of these `average' sources in an attempt to disentangle the AGN contribution to the FIR spectrum from the contribution from the cold dust associated with ongoing star formation.
	We then calculate SFRs from the fitted SEDs, and investigate the relationship between star formation and AGN power.
	From this analysis, we obtain the following results:

	1) In contrast to the study of \citet{page_suppression_2012}, we find no difference in the submm detection fraction with $L_{\mathrm{X}}$ (section~\ref{ssub:detections}); the results show no dependence of the fraction of sources detected on $L_{\mathrm{X}}$.
	Our study covers a significantly larger range in $L_{\mathrm{X}}$, and we do not find a cut-off in detections at $L_{\mathrm{X}} \approx 10^{44} {\mathrm{erg\: s}}^{-1}$ reported by \citet{page_suppression_2012} and \citet{barger_host_2015}.

	2) We find no decrease in stacked submm flux with increasing $L_{\mathrm{X}}$ (section~\ref{ssub:850_stack}).
	As a preliminary investigation, this again shows no indication of a decrease in SFR with increasing AGN power.

	3) SFRs calculated from the `average' SED fitting procedure, in which the contribution from warm and cold dust components to the FIR/submm SED are determined, also show no significant correlation with AGN luminosity (section~\ref{ssub:LAGN}).
	This adds to a body of recent work that finds no evidence of star formation being suppressed by powerful AGN \citep[e.g.][]{harrison_no_2012, stanley_mean_2017,suh_type_2017}.
	However, this may be due to the relative time-scales of AGN activity and star formation quenching, and our comparison of observational result to the \textsc{Horizon-AGN} hydrodynamical simulation suggests that this is not inconsistent with the scenario of AGN feedback quenching star formation in the galaxies that host them.

	In conclusion, by studying instantaneous X-ray luminosities and SFRs, we find no evidence that AGN activity affects star formation in host galaxies.

\section*{Acknowledgements}

	We thank Kristen Coppin, Elias Brinks, David Alexander and David Rosario for their insightful comments, and the anonymous referee for their useful suggestions. 
	JR and GM acknowledge support from the STFC [ST/N504105/1].
	JEG is supported by the Royal Society through a University Research Fellowship.
	The James Clerk Maxwell Telescope is operated by the East Asian Observatory on behalf of The National Astronomical Observatory of Japan; Academia Sinica Institute of Astronomy and Astrophysics; the Korea Astronomy and Space Science Institute; the Operation, Maintenance and Upgrading Fund for Astronomical Telescopes and Facility Instruments, budgeted from the Ministry of Finance (MOF) of China and administrated by the Chinese Academy of Sciences (CAS), as well as the National Key R\&D Program of China (No. 2017YFA0402700). Additional funding support is provided by the Science and Technology Facilities Council of the United Kingdom and participating universities in the United Kingdom and Canada.
	This research has made use of data from a number of sources, including the HerMES project (http://hermes.sussex.ac.uk/).
	HerMES is a \textit{Herschel} Key Programme utilising Guaranteed Time from the SPIRE instrument team, ESAC scientists and a mission scientist.
	\textit{Herschel} is an ESA space observatory with science instruments provided by European-led Principal Investigator consortia and with important participation from NASA.
	The HerMES data was accessed through the \textit{Herschel} Database in Marseille (HeDaM - http://hedam.lam.fr) operated by CeSAM and hosted by the Laboratoire d'Astrophysique de Marseille.
	We also use data from the Spitzer Space Telescope, which is operated by the Jet Propulsion Laboratory, California Institute of Technology under a contract with NASA, and both data and software provided by the High Energy Astrophysics Science Archive Research Center (HEASARC), which is a service of the Astrophysics Science Division at NASA/GSFC and the High Energy Astrophysics Division of the Smithsonian Astrophysical Observatory, as well as data obtained from the Chandra Data Archive and the Chandra Source Catalog.

	We made use of SciPy \citep{jones_scipy_2001}, NumPy \citep{van2011numpy} and  matplotlib, a Python library for publication quality graphics \citep{Hunter:2007}, as well as Astropy, a community-developed core Python package for Astronomy \citep{astropy}.




\bibliographystyle{mnras}
\bibliography{references} 




\bsp	
\label{lastpage}
\end{document}